\documentclass[1p]{elsarticle}

\usepackage{lineno,hyperref}
\usepackage{amsmath,amssymb,amscd}
\usepackage{graphicx}
\usepackage{comment}
\usepackage{enumitem}
\usepackage{dsfont}
\usepackage{physics}
\DeclareMathOperator{\TrG}{\widehat{G}}

\modulolinenumbers[5]

\journal{Journal of \LaTeX\ Templates}

\begin{document}

\begin{frontmatter}

\title{Psychophysical laws as reflection of mental space properties}

\author{Ihor Lubashevsky}
\address{University of Aizu, Ikki-machi, Aizu-Wakamatsu, Fukushima, 965-8580 Japan}
\ead{i-lubash@u-aizu.ac.jp}

\begin{abstract}
	
The paper is devoted to the relationship between psychophysics and physics of mind. The basic trends in psychophysics development are briefly discussed with special attention focused on Teghtsoonian's hypotheses. These hypotheses pose the concept of the universality of inner psychophysics and enable to speak about psychological space as an individual object with its own properties. 
Turning to the two-component description of human behavior (I. Lubashevsky, \textit{Physics of the Human Mind}, Springer, 2017) the notion of mental space is formulated and human perception of external stimuli is treated as the emergence of the corresponding images in the mental space. On one hand, these images are caused by external stimuli and their magnitude bears the information about the intensity of the corresponding stimuli. On the other hand, the individual structure of such images as well as their subsistence after emergence is determined only by the properties of mental space on its own. Finally, the mental operations of image comparison and their scaling are defined in a way allowing for the bounded capacity of human cognition. 
As demonstrated, the developed theory of stimulus perception is able to explain the basic regularities of psychophysics, e.g., (\textit{i}) the regression and range effects leading to the overestimation of weak stimuli and the underestimation of strong stimuli, (\textit{ii}) scalar variability (Weber's and Ekman' laws), and (\textit{iii}) the sequential (memory) effects.
As the final result, a solution to the Fechner-Stevens dilemma is proposed. This solution posits that Fechner's logarithmic law is not a consequences of Weber's law but stems from the interplay of uncertainty in evaluating stimulus intensities and the multi-step scaling required to overcome the stimulus incommensurability. 

\end{abstract}


\end{frontmatter}

\section{Introduction: From psychophysics to physics of mind}

The inner world of humans as an object admitting  analytical investigation  has attracted attention for many centuries starting from  ancient Greek philosophy. An important step in this direction was made in the second half of 19th century when in the frameworks of psychology and physiology a new branch of science---psychophysics---appeared. The subject-matter of this science may be characterized as the \textit{quantitative} description of the relationship between physical stimuli and the sensations and perceptions they produce.

Actually right at the birth of psychophysics it was recognized that there are \textit{outer} and \textit{inner} psychophysics partly overlapping. 
The former focuses the attention on the direct transformation of physical stimuli into human sensation of these stimuli and studies the stimulus-sensation relationship as a whole leaving aside possible mechanisms governing this transformation. Within the given approach a large amount of experimental data were accumulated in the 20th century. 

The latter brings into focus plausible mechanisms by which this transformation could be implemented, in particular, raises a question about their nature. Namely, whether the found laws of stimulus-sensation transformation are based on the particular physiology of sense organs or have to be attributed to the top level of mental processes. By the end of 20th century neuroscience of brain activity has provided evidence for the second standpoint, at least, it concerns the conscious component of human sensation related to \textit{evaluating} perceived stimuli.  

Attributing the mechanisms of psychophysical laws to the top level mental phenomena it becomes possible to analyze them from an alternative point of view. Following the gist of phenomenology developed in philosophy of mind in the 20th century these  laws might be elucidated ascribing some properties directly to the human mind as whole. In this case accepting the transformation of physical stimuli into the corresponding neural signals by the sense organs to be of a rather simple form, the psychophysical laws are explained turning to the complex behavior of the mind as an object with own properties and dynamics. 

The efficiency of such an approach turning to the top-down causation has been demonstrated by the recent development of synergetics---new science of self-organization---dealing with various non-linear phenomena in the inanimate world as well as in modeling brain activity, human behavior, and social phenomena. It is reflected in the appearance of novel disciplines such as mathematical psychology including its applications to psychophysical problems, socio-physics, and econo-physics. 

In order to describe the behavior of humans affected by external stimuli and responding to them I put forward a general approach that can be referred to as \textit{physics of the human mind} \cite{ihor2017bookmind}. Its pivot point is the introduction of two complementary components---objective and subjective ones. The objective component is related to the physical world and governed by the classical laws of physics. Returning to psychophysical issues the sense organs and their functioning in transformation of physical stimuli into neural signals should be regarded the objective component. In this case the subjective component represents the mind-body interaction from the standpoint of mind. The mind is treated as an object with its own properties. These properties reflect neurophysiological processes in the brain but not are enslaved by them. In this way we gain the ability to allow for the basic human properties in the corresponding mathematical models appealing to our consciousness as experienced from the first-person point of view.  

The purpose of the present paper is to demonstrate that within physics of mind it is possible to construct a rather simple model of how perceived stimuli are reflected in the mind and how the mind operates with them. Base on a few premises this model reproduces many psychological laws seeming to be mutually independent and proposes a solution to some challenging problems in psychophysics.

\section{Basic psychophysical laws \& Fechner-Stevens dilemma}

Psychophysics is rooted in experimental psychology of the late 19th century, its birth is related to the publication of the famous book  \textit{Elements of Psychophysics} (1860) by Gustav Theodor Fechner (1801--1887). In this book Fechner posed a number of questions about the interrelation between the intensity of various physical stimuli and human sensation of these stimuli (for introduction see, e.g., \cite{gescheider1997psychophysics,Grondin2016psychologypsy}). Actually at birth psychophysics underwent bifurcation, Fechner himself drew a crucial distinction between \textit{outer} and \textit{inner} psychophysics, which differ in subject-matter; a detailed review of this issue can be found in Refs.~\cite{scheerer1992fechner,robinson2010fechner}. According to Fechner, human sensation is implemented as a sequence of two transformations with own properties: 
\begin{itemize}[topsep=0.15\baselineskip, itemsep=0.1\baselineskip,  parsep=0.15\baselineskip]
	\item \textit{stimulus} ${}\rightarrow{}$ \textit{psychophysical processes},\footnote{In modern terms psychophysical process may be treated as the generation of neural signal by the corresponding sense organ.} 
	\item \textit{psychophysical processes} ${}\rightarrow{}$ \textit{sensation}.
\end{itemize}
Outer psychophysics deals with the two transformations as a whole and studies the stimulus-sensation relationship based on direct observation and experimentation. 

Inner psychophysics focuses on the latter transformation and has to draw inferences from the findings of outer psychophysics. Experimentation and quantification are rather problematic in inner psychophysics; for this reason inner psychophysics was rejected for a long time starting from Fechner's contemporaries. Nevertheless, Fechner himself considered that regularities found in outer psychophysics and being only approximate become strictly valid in inner psychophysics, which makes inner psychophysics important branch of science on its own. For the modern state of the art in inner psychophysics taking into account the advances in cognitive neuroscience analyzing the mapping between mental quantities and the brain states a reader may be referred to \cite{fnins.2016.00190}. 

Based on the findings of his mentor Ernst Heinrich Weber (1795--1878), Fechner formulated one of the basic psychophysical laws, \textit{Weber's law}, which is widely accepted to hold for all our sensory modalities. 
Weber's law relates the perception threshold $\Delta S$ of variations (the just-noticeable difference) in the intensity $S$ of some physical stimulus, e.g., sound or light, to its intensity $S$ via the proportionality with a coefficient $\kappa_w$ often referred to as the Weber fraction,
\begin{equation}\label{Weberlaw}
\Delta S = \kappa_w S \ \text{or its later modification} \ \Delta S = \kappa_w S + \eta
\end{equation}     
allowing for this stimulus of low intensity via adding some constant $\eta$ interpreted as the result of sensory noise. For example, the characteristic values of Weber fractions for visual length of lines, brightness, and loudness are estimated as $\kappa_w = 0.04$, 0.08, and 0.1, respectively \cite{bairdnoma1978}. 

Appealing to Weber's law, Fechner put forward the principle that an arithmetic series of mental magnitudes of perceived physical stimuli should correspond to a geometric series of their intensities (energies) \cite[e.g.,][]{gescheider1997psychophysics}. In the framework of this principle the ratio $\Delta S / S$ is regarded as the induced increment in the sensory magnitude $\Delta M$ of the given physical stimulus, whence it follows that 
\begin{equation}\label{Fechnerlaw}
M =  \kappa_w \ln S \quad \text{for $S \gg \eta/\kappa_w$}.
\end{equation}     
This logarithmic relationship is often referred to as \textit{Fechner's law}. It should be emphasized that the introduction of the additive term in Weber's law \eqref{Weberlaw} eliminates the formal logarithmic singularity of Fechner's law at $S\to 0$. An alternative approach to tackling this singularity was proposed by Guilford~\cite{guilford1932generalized} assuming Weber's law to be of the form $\Delta S \propto S^{d}$ with the exponent $d<1$ meeting the condition $1-d \ll 1$.

Fechner's description of the stimulus-sensation relationship was dominating for a hundred years until Stanley Smith Stevens (1906--1973) put forward a new psychophysical law \cite{stevens1957psychophysical,Stevens1961Science} called after him \textit{Stevens' law}. When subjects are asked to \textit{quantify } (numerically or, e.g., via squeezing a handgrip dynamometer) their sense of subjective intensity the $M(S)$-relationship  is described more effectively by a power law
\begin{equation}\label{Stevenslaw}
M  = k S^{\beta_s},
\end{equation}
where the coefficient $k$ and the exponent $\beta_s$ are some constants determined by the physiological mechanisms of the corresponding sensory modality.  The Stevens exponent $\beta_s$ can be less or larger than or equal to 1, e.g., for visual length of lines, brightness, and loudness $\beta_s = 1.0$, 0.5, and 0.67, respectively, for electric shock $\beta_s = 3.5$ \cite[e.g.,][]{gescheider1997psychophysics}. 

The appearance of two different laws competing for the description of the stimulus-sensation relationship gave rise to long-term debates in psychophysics including the present time. These debates are focused on the  possibility of reconciling Fechner's law and Steven's law because both of them being the vary basic laws of psychophysic, on one hand,  concern quantifying the inner sensation  of external physical stimuli and, on the other hand, have functionally different forms; I call this issue the \textit{Fechner-Stevens dilemma} for short. Broadly speaking, the subject of these debates may be interpreted as to whether it is possible that the two forms of the relationship $M = M(S)$ given by Fechner's law and Stevens' law are really mathematically equivalent over some range of external stimulus intensity \cite[][Chap.~3]{Norwich1993weberK}. It should be noted that this dilemma can be also posed at a high level of describing the human mind in dealing with cultural aspects of human society \cite{Dehaene1217,Cantlon38}.

Almost immediately after Stevens' publication, MacKay \cite{MacKay1213} proposed the concept of the relationship between the intensity $S$ of physical stimulus and the perceived magnitude $M$ (the internal activity) via the neural network in order to reconcile the two laws with each other. Namely, accepting the neuron operating frequencies caused by the stimulus intensity and the brain internal activity to be related to $S$ and $M$ via Fechner's logarithmic law and the stimulus-sensation comparison to be implemented via a linear relationship of the two frequencies, MacKay comes directly to Stevens' law~\eqref{Stevenslaw}. Similar ideas were considered also in \cite{treisman1964,ekman1964power}, in addition see \cite{Ekman1965Scaling,Cook1967powerlaw} for their discussion.
Nowadays this concept has developed into a branch of science that could be called neural psychophysics focusing its attention on neural mechanisms governing human inner perception of physical stimulus \cite[for a review see, e.g.,][and references therein]{Johnson2002neraul,PhysRevE.65.060901,billock2011honor,journal.pcbi.1003781}.

Another approach to tackling the Fechner-Stevens dilemma is to modify Weber's law in the vicinity of some special points accompanied by additional modifications in the understanding of how the internal sense of physical stimuli is formed with the increase in the stimulus intensity. The discussion of this approach was induced by publication \cite{krueger1989reconciling} followed by a large number of commentaries published in the same journal; for a review see, e.g., \cite{Laming1997BookMeasurement,Norwich1997,Zwislocki2009}.


The crucial point in the psychophysics development playing an essential role in my further constructions is related to the idea of converting the description of stimulus-sensation regularities from outer psychophysics to inner psychophysics posed by G\"osta Ekman (1920--1971) \cite{ekman1959webers}. Instead of dealing with the intensity of physical stimuli he proposed to analyze the basic laws of human sensation in terms of sensory magnitudes. In particular, he demonstrated that the relationship between the variation of magnitude $\Delta M$ matching the just-noticeable difference $\Delta S$ of the corresponding physical stimulus and the sensory magnitude obeys the law 
\begin{equation}\label{Ekmanlaw}
\Delta M = \kappa_e M
\end{equation}
similar to Weber's law~\eqref{Weberlaw} in form. Relationship~\eqref{Ekmanlaw} is often referred to as \textit{Ekman's law} and below the coefficient $\kappa_e$ will be called Ekman fraction. It should be noted that Ekman's law as a conjecture was put forward by Franz Brentano (1838--1917) whose work influenced also the development of phenomenology as a branch of philosophy \cite{brentano1874}. 

Keeping in mind my further constructions I want to emphasize that Ekman's law:
\begin{itemize}[topsep=0.15\baselineskip, itemsep=0.1\baselineskip,  parsep=0.15\baselineskip]
	\item  presents a series obstacle to Fechner's logarithmic interpretation of the stimulus-sensation relationship based on Weber's law,
	\item is regarded as an immediate consequence of Weber's law and Stevens's law \cite[e.g.,][]{Laming1997BookMeasurement}.
\end{itemize}
As far as the latter item is concerned, the three laws are, in fact, not mutually independent. Weber's law admits much more reliable way of verification in comparison with Ekman's law. Nevertheless, below I will regard Ekman's law to be the basic one whereas Weber's law to be a  consequence because Ekman's law can be explicitly attributed to inner psychophysics directly dealing with the mind. 

\section{Universality of inner psychophysics}\label{Sec:universality}

The model to be developed turns to Ekman's law as well as a number of hypotheses about the universality of inner psychophysics put forward by Robert Teghtsoonian (1932--2017). Based on the experimental data summarized, in particular, by Poulton~\cite{Poulton1967,poulton1968new} and using the notion of  \textit{dynamic range}---the span (usually given in $\log$-units) from the lowest to highest stimulus intensities or sensory magnitudes admitting a continuous grading---Teghtsoonian claims that \cite{teghtsoonian1971exponents,Teghtsoonian1973range,Teghtsoonian1978}:
\begin{enumerate}[topsep=0.15\baselineskip, itemsep=0.1\baselineskip,  parsep=0.15\baselineskip]
	
	\item\label{Th:1} The Ekman fraction, in contrast to the Weber fraction, is approximately the same value equal to $\kappa_e= 0.03$ for the main sensory modalities, which assumes the strict relationship $\beta_s= \beta_s(\kappa_w)$ between the Weber fraction and the Stevens exponent.
	
	\item\label{Th:2} A wide variety of perceptual continua are characterized by a common maximal dynamic range $R_M$ of sensory magnitude with the estimate $\log R_M\sim 1.5$. 
	
	\item\label{Th:3} \textit{Crossmodal property}. The maximal dynamic range $R_S$ of stimulus intensity and the Stevens exponent $\beta_s$ are related such that the product $\beta_s\cdot\log R_S=\log R_M$ is an approximately constant value.
	
	\item \label{Th:4} \textit{Intramodal properties}. In quantifying two stimuli within a given perceptual continuum the ratio $r_S$ of their intensities and the ratio $r_M$ of their sensory magnitudes are related via the expression 	$\log r_M = \beta_s\cdot \log r_S + k$ where the constant $k\sim 0.1$. Broadly speaking, humans, on the average, tend to constrict the subjective dynamic range in quantifying physical stimuli. In particular: 
	\begin{itemize}
		\item Within a fixed range of test stimuli the magnitude estimates are systematically biased towards the center of the tested range as it has been figured out and studied previously by Hollingworth~\cite{Holligh1910central} and Stevens \& Greenbaum~\cite{Stevens1966}. This bias called the \textit{regression effect} causes an underestimation of large magnitudes and an overestimation of small ones. The underestimation and overestimations of long and short time intervals, respectively, was recognized even earlier by Karl von Vierordt (1818--1884), which is reflected in his famous book \textit{The Experimental Study of the Time Sense} (English translation) (1868) \cite[e.g.,][for details]{10.1080/09541440802453006}.
		
		\item This bias towards the center becomes more pronounced with increasing magnitudes, which is called the \textit{range effect} \cite{Teghtsoonian1973range,Teghtsoonian1978}.
	\end{itemize}
\end{enumerate}
To complete this description of the basic psychophysical properties I have added the following item accumulating the results of other scholars. 
\begin{enumerate}[topsep=0.15\baselineskip, itemsep=0.1\baselineskip,  parsep=0.15\baselineskip]
		\item[5.]\label{Th:55} Judgments on perceived magnitudes are also affected by the recent  history of evaluating physical stimuli belonging to the same perceptual continuum or, even, deferent one, which is often called \textit{sequential effects}. Their various aspects were thoroughly studied in the last century and for a recent review a reader may be referred, e.g., to \cite{wrap2009,matthews2009inter,Podlesek2010,fpsyg.2017.00933}. In particular, the sequential effects give rise to \textit{hysteresis} in stimulus evaluation. It means that the sensory magnitude ascribed to a given stimulus by an observer can be different when the given stimulus is approaches from stimuli of higher or weaker intensity \cite{stevens1957psychophysical,Cross1973}. 
\end{enumerate}
As far as Proposition~\ref{Th:1} is concerned with, the analysis of a wider collection of experimental data by Laming \cite{Laming1997BookMeasurement} demonstrates that the Ekman fraction obtained in the way used by Teghtsoonian \cite{teghtsoonian1971exponents} can deviate substantially from the value  $\kappa_e= 0.03$. It poses doubts about the existence of a strict relationship between Weber's law of \textit{sensory discrimination} and Stevens's law of \textit{sensory judgment} \cite{Laming1997BookMeasurement}. Below I argue for its existence and relate the scattering of calculated values of the Ekman fraction to the use of data obtained at the boundary of judgment possibility.

The aforementioned properties enabled Teghtsoonian~\cite{teghtsoonian1971exponents,Teghtsoonian1973range,amerjpsyc.125.2.0165}  to posit a single central macrolevel mechanism responsible for all judgments of sensory magnitude. Within this mechanisms the role of various receptor systems is reduced to performing the necessary expansions or compressions required to map the widely varying dynamic ranges of physical stimuli into the constant range of subjective magnitudes. The similarity between human perception of physical stimuli based on our sense organs and the mental evaluation of abstract objects like numbers  \cite{Baird1975NumI,Noma1975NumII,Weissmann1975,Baird1975NumIV,Baird1975NumV} is an additional argument for this mechanism.

The data accumulated in fMRI, EEG, and neuropsychological investigations also argue for this central mechanism. In particular Walsh and Bueti~\cite{walsh2003atom,Bueti1831} put forward a new paradigm about human judgment and evaluation of external stimuli called ATOM (``A Theory Of Magnitude''). The ATOM supposes that various dimensions of magnitude information, such as space, time, and quantity, are encoded by ``common neural metrics'' in the parietal cortex, which explains the emergence of common \textit{neurocognitive} mechanisms governing human perception of various physical stimuli. Naturally, the situation can be more complex \cite{PINEL2004983}. For a review of arguments for and against this universality a reader may be referred to \cite{COHENKADOSH2008132,dormanpesenti2012,fnhum.2016.00500}. Moreover Hayes \textsl{et al.} \cite{fpsyg.2014.00529} generalized the ATOM by extending its scope onto memory, reasoning, and categorization traditionally treated as separate components of human cognition. The exemplar-based account of the relationship between categorization and recognition \cite[][]{nosofsky2012activation} is also rather close to this paradigm.

The following premises accepted nowadays in psychophysic also underlies the concept of central mechanism of stimulus evaluation. \textit{First}, the stimulus perception is considered to be implemented into two steps: the conversion of physical stimuli into neural signals by sense organs (perceptual subsystem) and the mental evaluation of these signals based on the central subsystem functioning. After Fechner, Treisman \cite{treisman1964} seems to be the first who supposes the perceptual and central subsystems to be governed by their \textit{own} laws. Naturally, the two subsystems mutually affect each other \cite[see, e.g.,][for a review and the discussion of similar concepts]{Baird1997}.  This type description, in particular, was proposed in  \cite{petrov2003addmult,petrov2005dynamics} where the role of human memory is accentuated. 

The \textit{second} premise focuses attention on the cognitive evaluation of perceived stimuli where the introduction of \textit{psychological space} takes the central place. The concept of psychological space is rooted in the works of Louis Leon Thurstone (1887--1955) formulated the \textit{law of comparative judgment} converted then into the \textit{law of categorical judgment} by Warren S. Torgerson (1924--1999).
However the psychological space as an object admitting sophisticated mathematical description appears in the works by Robert Duncan Luce (1925--2012) \cite[e.g.,][also \cite{luce2013} for a general discussion]{luce1959possible,luceMathBio1963.ch3,luceMathBio1963.ch5,luceMathBio1965.ch19}. The gist of these constructions is the use of the probabilistic properties of human decision-making in categorizing various external stimuli for constructing the topology of the inner psychological space, whose points are various categories.  Exactly the introduction of psychological space with own properties and governing the final step in stimulus evaluation explains the universality of psychophysical laws for all the sensory modalities \cite[e.g.,][]{Shepard1957,shepard1958stimulus.2,shepard1958stimulus.1,shept1700004science,Nosofsky1992}. Among the recent publications this universality has been also discussed, e.g., in \cite{DZHAFAROV20051,DZHAFAROV2005125} and \cite{NosofkyPair2015random,WINTER2015209b,WINTER2015209a}. In this connection I want to note the Bayesian approach to describing the interaction between the perceptual and central subsystems as well as their individual operations turning to the formalism of conditional probabilities \cite[][for review]{Petzschner17220,PETZSCHNER2015285}.  The psychological space can be also introduced axiomatically  \cite[e.g.,][]{NARENS1996109,Narens2002book}.

The \textit{third} premise concerns the nonlinear properties of inner representation of external stimuli. In particular, it posits that the uncertainty in evaluating sensory magnitudes is determined mainly by the functioning of the central subsystem rather than the properties of noise attributed to the corresponding physical stimulus and the physiological transformation of physical effects into neural signals by sense organs. Petrov and Anderson~\cite{petrov2003addmult,petrov2005dynamics} allow for this nonlinearity endowing the relationship between the sensory magnitude $M$ and the physical stimulus intensity $S$ with an additive noise whose amplitude depends on the sensory magnitude itself. Their model is based on the interpretation of Weber's law \eqref{Weberlaw} by Ekman~\cite{ekman1959webers} which considers the just-noticeable differences $\Delta S$ to be caused by perceptual noise additively entering the relationship $S\mapsto M$ with amplitude proportional to the mean value of sensory magnitude, 
\begin{equation}
\langle (\delta M)^2 \rangle^{1/2} = \kappa_e \langle M\rangle\,,
\end{equation}
which is probabilistic interpretation of Ekman's law~\eqref{Ekmanlaw}.  Petrov's model accepts the relationship between the sensory magnitude and physical stimulus to be of the form
\begin{equation}\label{Petrov}
M = kS^{\beta_s}(1+ \kappa_e \xi)\,.
\end{equation}
Here $\xi$ is a noise of unit amplitude. Assuming the noise contribution to the sensory magnitude to be small and the just-noticeable differences $\Delta S$ to be specified by the amplitude of the random fluctuations in the sensory magnitude, Exp.~\eqref{Petrov} immediately gives rise to Weber's law \eqref{Weberlaw} with $\kappa_w\sim\kappa_e/\beta_s$. The probabilistic concepts turning, in particular, to the notion of entropy have been used to model psychophysical regularities \cite[e.g.,][]{Norwich1993weberK,Norwich1987,Norwich2005,Nizami2009}.

\section{Goal of the present work}

The psychological space as an object with own properties attributed to it as whole may be regarded as one of the basic elements appearing in the framework of the philosophical approach to describing the human mind called \textit{phenomenology}. Phenomenology was launched in the first half of the 20th century by Edmund Husserl, Martin Heidegger, \textit{et al.} Broadly speaking, phenomenology studies structures of conscious experience as experienced from the \textit{first-person point of view} \cite[e.g.,][for introduction]{sep-phenomenology,galenda2012phenomenological}. In book~\cite{ihor2017bookmind} following the gist of phenomenology I proposed a two-component approach to describing human behavior and actions. This approach introduces two \textit{complementary} components, \textit{objective} and \textit{subjective} ones. The objective component represents the physical (outer) world governed by the classic laws of physics. The subjective component represents our inner world and is assumed to be governed by its own laws irreducible to the laws of the objective components within the given phenomenological approach. For this reason I claim that we need an \textit{individual} mathematical formalism for describing mental phenomena in terms that can be formulated turning to the first-person experience. As a validation of the given approach efficiency, we may consider its feasibility of constructing mathematical models for particular phenomena that (\textit{i}) can provide us with complete, self-consistent description, (\textit{ii}) turn in their construction to the basic properties of our mind, and (\textit{iii}) minimize the number of accepted assumptions and introduced mathematical notions.   

In the present paper I develop a phenomenological model for human evaluation of external physical stimuli that from a rather general standpoint explains the universality of inner psychophysics and proposes a solution to the Fechner-Stevens dilemma. In what follows I will confine my consideration to the \textit{prothetic} (in Stevens's terminology \cite{stevens1957psychophysical}) perceptual continua such as loudness and brightness admitting continuous variation in the intensity of the corresponding stimuli without changes in their categorization. The \textit{metathetic} perceptual continua, e.g., color or pitch require an individual consideration.

Before passing directly to the model description it should be noted that the notion of \textit{mental space} takes the central place in the further constructions. It is similar to the concept of psychological space discussed above. Nevertheless I prefer to use the term mental space to emphasize that its elements---mental images of some physical objects---are irreducible to their physical sources.  Broadly speaking, within the two-component description these mental images are entities of the subjective component. 

\section{Model}

\subsection{General comments}

\begin{figure}[t]
	\begin{center}
		\includegraphics[width=0.65\columnwidth]{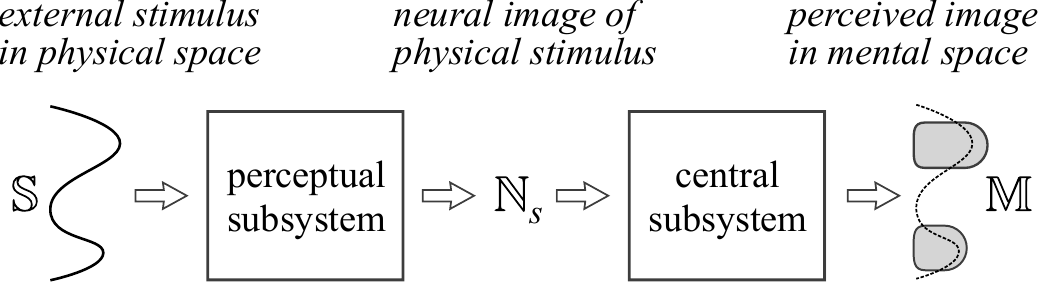}
	\end{center}
	\caption{The two-step process of stimulus perception with the intermediate neural image $\mathbb{N}_s$ of an external (physical) stimulus $\mathbb{S}$.}
	\label{F:SNM}
\end{figure}

As the pivot point of the proposed concept, perception of a physical stimulus is assumed to be a two-step process illustrated in Fig.~\ref{F:SNM}. At the first step an external (physical) stimulus $\mathbb{S}$ is converted by the corresponding sense organ (perceptual subsystem) into a neural signal $\mathbb{N}_s$ which can be analyzed by the central subsystem. At the second step the central subsystem converts this neural signal into an image $\mathbb{M}$---an element of the mental space. 

To avoid possible misunderstanding and expecting critical comments from reductive physicalists I note the following. From the standpoint of reductive physicalism,\footnote{Reductive physicalism is a philosophical account of human mind accepting that all the objects as well as their properties are reducible to the underlying physical states and processes. As a result, mental states and processes cannot have causal power which may be attributed only to physiological, in particular, neural processes in the brain \cite[see, e.g.,][]{sep-physicalism}.} only neural processes appear at the basic level of description of stimulus perception. This feature is reflected in the standard presentation of composite perception process: (\textit{i}) the conversion of physical stimuli into neural signals proceeded by the central subsystem and then (\textit{ii}) human response $\mathbb{R}$ caused by these signals, $\mathbb{S}\mapsto\mathbb{N}\mapsto\mathbb{R}$ \cite[e.g.,][]{petrov2003addmult,petrov2005dynamics}.

The main  difference between the standard approach and my approach is that I consider separately the neural signal $\mathbb{N}_s$ generated by sense organs (the perceptual subsystem) and, then, I introduce the mental space reflecting the brain activity responsible for \textit{conscious} evaluation of perceived stimuli.


According to reductive physicalism an epiphenomenal approach to describing human perception of physical stimuli cannot be accepted as fundamental because it operates with objects having no causal power. Therefore models directly dealing with mental images of physical stimuli cannot be used for explaining the basic properties of psychophysical laws.  

Nevertheless, numerous examples from physics, Landau's phenomenological theory of phase transitions is one of them, demonstrate that theories dealing with, strictly speaking, epiphenomenal entities as objects \textit{affecting directly}  the behavior of analyzed systems can be highly efficient. The model I propose belongs to the class of such phenomenological theories. 

The basic advantage of the given approach is the possibility of ascribing to elements of mental space individual properties existing on their own. In terms of two-component description \cite{ihor2017bookmind}
\begin{itemize}[topsep=0.15\baselineskip, itemsep=0.1\baselineskip,  parsep=0.15\baselineskip]
\item[-] physical stimuli are entities of the objective component, 
\item[-] the mental space with images caused by physical stimuli is a particular instantiation of the subjective component, and 
\item[-] the neural signals directly generated by sense organs determine the interaction of the two components.    
\end{itemize}
In what follows I will analyze stimulus perception described in terms of the relationship of the stimulus intensity $S$ and the perceived magnitude $M$. The first quantity, $S$, is ascribed to the analyzed physical stimulus, the second one, $M$, is attributed to the image of this stimulus in the mental space.  The neural signal of intensity $N$ generated by the corresponding sense organ responding to the stimulus $\mathbb{S}$ is considered, on one side, to be completely determined by the external stimulus, $N= N(S)$, and, on the other side, to play the role of a source for the image $\mathbb{M}$ emerging in the mental space. The main attention will be focused on the emergence of such images, whereas the $N(S)$-relationship merely reflecting the underlying physiological processes is specified by the ansatz 
\begin{equation}\label{M:StevPh}
N = A_p \cdot S^{\beta_{p}}
\end{equation} 
where $A_p$ and $\beta_p$ are some constants. Relationship~\eqref{M:StevPh} may be regarded as the physiological component of Steven's law. Its nonlinearity (for $\beta_p\neq1$) allows for the possibility of the perceptual system adaptation to the specific conditions of environment, in particular, to the intensity of external stimuli \cite[see, e.g.,][for the discussion of physiological mechanisms of various sensory modalities from the standpoint of psychophysics]{Grondin2016psychologypsy}. When there are no special sense organs perceiving a give type of physical stimuli  the neural signal $\mathbb{N}$ can be generated by some internal mechanism. In particular, it is the case for time perception and its implementation can be based on internal physiological ``clocks'' \cite[e.g.,][for an introduction to the perception of time]{Wearden2016psychology}. This situation is also allowed for by expression~\eqref{M:StevPh} with $\beta_{p}=1$, i.e., where the physical stimulus intensity $S$ and the amplitude $N$ of neural signals may be identified within some proportionality. For time perception the number of internal clock ticks and the duration of time interval exemplify this proportionality.

\subsection{Mental space}

The image $\mathbb{M}$ emerging in the mental space after processing a certain neural signal $\mathbb{N}$ is the main object to be discussed in the remaining part of the present paper. It is assumed to possess the following properties.
\begin{itemize}
	\item[P1:] The image $\mathbb{M}$ bears a certain information about the amplitude $N$ of signal $\mathbb{N}$. It does not mean that this information is coded in units specifying various physiological features of signal $\mathbb{N}$. The form of coding may be rather abstract and rid of many particular details characterizing the spatio-temporal structure of the neural signal $\mathbb{N}$. 
	For this reason the image $\mathbb{M}$ is assumed to possess some property called \textit{magnitude} representing in some way the amplitude $N$ of neural signal. Naturally there could be other aspects of the image $\mathbb{M}$ representing other features of neural signal, i.e., the information about sensory modality.
	
	\item[P2:] After the emergence caused by the neural signal $\mathbb{N}$, the image $\mathbb{M}$ no longer needs the signal $\mathbb{N}$ to be present. So mental images of physical stimuli are individual objects existing on their own in the mental space, which implies the memory-based nature of the mental space. The introduction of stimulus images in this way opens a gate to modeling their own dynamics in the memory.
	
	\item[P3:] Due to our consideration being confined to a certain \textit{prothetic} perceptual continuum (e.g., loudness) the corresponding images admit direct comparison with one another based only on their magnitudes because other aspects of these images (e.g., sound frequency) reflecting the nature of the given perceptual continuum and the corresponding sensory modality can be the same.

\begin{figure}
	\begin{center}
		\includegraphics[width=0.85\columnwidth]{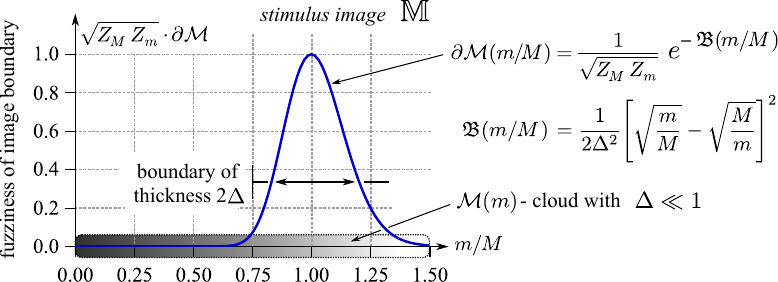}
	\end{center}
	\caption{The mental image $\mathbb{M}$ with magnitude $M$ as the $\mathcal{M}(m)$-cloud---a fuzzy segment with blurred boundary on the $m$-axis. The fuzziness of the image boundary $\partial\mathcal{M}(n/M,\Delta)$ is specified by function~\eqref{A1:1}, in this illustration the parameter $\Delta = 1/8$ corresponding to the boundary thickness of 25\% was used.}
	\label{Fig:Cloud}
\end{figure}

	\item[P4:] The image $\mathbb{M}$ possesses a certain \textit{universal} structure represented in Fig.~\ref{Fig:Cloud}.
	 On the axis of magnitudes (further the $m$-axis) the image $\mathbb{M}$ is represented as some fuzzy region---the $\mathcal{M}(m)$-cloud---with blurred boundary of thickness $\Delta$ centered at a point $M$. The distance between the $m$-axis origin and the latter point is referred to as the magnitude $M$ of the image $\mathbb{M}$.
		
	\quad The $\mathcal{M}(m)$-clouds is characterized by two parameters: the image magnitude $M$ and the relative thickness  $\Delta$ of its boundary $\partial\mathcal{M}$ specified as a function $\partial\mathcal{M}(m/M, \Delta)$ except for possible cofactors arising via integration over the $m$-axis under certain conditions. 
	Besides, keeping in mind the general features of this type blur and the further use of the given function in the image comparison I accept two additional assumptions.
\begin{itemize}
	\item The function $\partial\mathcal{M}(m/M, \Delta)$ is supposed to be symmetric with respect to the argument exchange $m\leftrightarrow M$, i.e., $m/M \leftrightarrow M/m$.
	\item The relative thickness $\Delta$ of the boundary $\partial\mathcal{M}$ is considered to be a small parameter, $\Delta \ll 1$. So in further mathematical manipulations I may confine them to the leading order in $\Delta$.
\end{itemize}  

Turning to the general properties of the function $\partial\mathcal{M}(m/M, \Delta)$ attaining its maximum at $m= M$ the following ansatz 
\begin{gather}\label{A1:1}
\partial \mathcal{M}(m/M,\Delta) = \frac1{\sqrt{Z(M\Delta)\, Z(m\Delta)}}\exp\Big\{-\mathfrak{B}(m/M,\Delta)\Big\},\\
\intertext{where}
\label{A1:2}
\mathfrak{B}(m/M,\Delta) = \frac1{2 \Delta^2} \frac{(m-M)^2}{mM} = \frac1{2 \Delta^2}
\Bigg[\sqrt{\frac{m}{M}} - \sqrt{\frac{M}{m}} \Bigg]^2   
\end{gather}
is accepted for the boundary fuzziness. 
The cofactors $\{Z(\ldots)\}$ are specified by the equality 
\begin{equation}\label{A1:3}
\bra{\mathbb{M}}\ket{\mathbb{M}} \stackrel{\text{def}}{{}={}} \int\limits_0^\infty [\partial \mathcal M(m/M,\Delta)]^2 dm = 1\,,
\end{equation}
imposed on the image $\mathbb{M}$. Whence it follows that in the leading order in $\Delta$
\begin{equation}\label{A1:4}
Z(M\Delta) = \big( \sqrt\pi M\Delta\big)^{1/2}\,, \quad Z(m\Delta) = \big( \sqrt\pi m\Delta\big)^{1/2}
\end{equation}
Leaping ahead, I note that equality~\eqref{A1:3} admits interpretation such as ``any image always coincides with itself.'' 
		
	\item[P5:] Strictly speaking, when we deal with different images in the mental space the magnitude $M$ of image $\mathbb{M}$ cannot be used as just a number to analyze their interrelations \textit{directly}. Nevertheless, attributing magnitudes to these images should enable us to introduce the following two general operations implemented consciously in the mind: 
	\begin{itemize}
		\item[\textit{a})] A \textit{qualitative comparison} of two images $\mathbb{M}_1$ and $\mathbb{M}_2$ with respect to which one has a smaller or larger magnitude, or whether their magnitudes are identical, i.e.,\footnote{Strictly speaking, the inference that a given pair of images are incommensurable \textit{in principle} may be also introduced as a plausible result of qualitative comparison. However this generalization desires an individual consideration.}  
		\begin{equation}\label{Mod:P3a}
			\mathbb{M}_1\prec \mathbb{M}_2\,, \quad \mathbb{M}_1 \succ \mathbb{M}_2\,, \quad \mathbb{M}_1\simeq \mathbb{M}_2\,.
		\end{equation}
		I have used the symbols ``$\prec$,'' ``$\succ$,'' and ``$\simeq$'' to denote these interrelations instead of ``$<$,'' ``$>$,'' and ``$=$'' in order to avoid a possible confusion between the introduced relations and the standard relations of real numbers.    
		
		\item[\textit{b})] A certain scaling  of the image amplitudes to be called the \textit{image fusion} determined by the structure of these images
		\begin{equation}\label{Mod:P3b}
		 n \otimes\mathbb{M}_1\mapsto \mathbb{M}_n \ \text{ where $n$ is an integer.} 
		\end{equation} 	 
	\end{itemize}     
\end{itemize}    
It is worthy of noting that the possibility of ascribing such properties to results of mental operations with perceived physical stimuli is due to the introduction of mental space as some individual object with its own properties. 


\subsection{Comparison of adjacent images: Ekman's law}\label{A:Sec:2.1}

When two physical stimuli affect the perceptual subsystem simultaneously there could be a certain physiological mechanism enabling their comparison with higher accuracy than, e.g., in the case when these stimuli are separated by some time gap and their comparison requires memory. 

In the case of simultaneous stimuli their images $\mathbb{M}_1$ and $\mathbb{M}_2$  are supposed to be adjacent to each other in the mental space and the arguments $m_1$ and $m_2$ of their clouds may be identified. 
It enables us to calculate $\bra{\mathbb{M}_1}\ket{\mathbb{M}_2}$ following the standard rule:
\begin{equation}\label{A2:3}
\bra{\mathbb{M}_1}\ket{\mathbb{M}_2} = \int\limits_0^\infty \partial M_1(m)\partial M_2(m)\, dm
\end{equation}
and quantify the proximity of the two images using this value. According to the premise~P4 both the clouds $\mathcal{M}_1(m)$,  $\mathcal{M}_2(m)$ should possess the same structure with identical relative thickness of their boundaries, $\Delta_1=\Delta_2\stackrel{\text{def}}{{}={}}\Delta_e/\sqrt2$. Indeed, at the moment of stimulus action, on one hand, the $\mathcal{M}(m)$-cloud formation has to be enslaved by the physical stimulus and, on the other hand, its form is determined by the properties of mental space. The two statements can be reconciled with each other assuming that the physical stimuli determine the magnitudes of their images whereas the form of $\mathcal{M}(m)$-clouds is determined by the properties of mental space that are not affected by memory effects and reflect only the \textit{current} neural signals generated by the perceptual subsystem.      
Direct calculations based on Exp.~\eqref{A1:1} within the leading order in $\Delta_e$ yield
\begin{equation}\label{A1:12}
\bra{\mathbb{M}_1}\ket{\mathbb{M}_2} = \exp\big\{-\mathfrak{B}(M_1/M_2,\Delta_e) \big\}\,.
\end{equation}
In probabilistic terms this expression admits the interpretation as the probability of classifying the images  $\mathbb{M}_1(m)$,  $\mathbb{M}_2(m)$ as equivalent in magnitude and the function $\mathfrak{B}(M_1/M_2,\Delta_e)$ may be regarded as the potential measure of image proximity or just the \textit{proximity potential} for short. 

The obtained result allows us also to estimate the just noticeable difference in the image magnitudes of \textit{simultaneous} stimuli, $\delta M$,  setting 
\begin{equation*}
\mathfrak{B}\big[(M+\delta M)/M,\Delta_e\big] \lesssim 1\,.
\end{equation*}    
Whence it follows that 
\begin{equation}\label{A2:5}
\frac{\delta M}{M} \stackrel{\text{def}}{{}={}} \kappa_e \sim \Delta_e\,,
\end{equation}    
which is Ekman's law. The available experimental data for the Ekman fraction $\kappa_e\approx 0.03$ discussed above give the numerical estimate for the parameter $\Delta_e$ in the case of currently active stimuli. 

The obtained expressions for the proximity of two images in magnitude illustrate the basic formalism to be developed for more general situations.

\subsection{Memory-based image comparison}

Mental operations with images, e.g., creating mental copies of an image $\mathbb{M}$, retaining an active stimulus image in memory and retrieving it later changes the \textit{states} of these images and thus can affect both the $\mathcal{M}(n)$-cloud  characteristics, the magnitude $M$ and the relative measure $\Delta$ of cloud blur. Unconscious change in the magnitude $M$ should be regarded as some bias, which lies outside the present analysis. Here I focus our attention on changes in $\Delta$. 

\begin{figure}
	\begin{center}
		\includegraphics[width=0.5\columnwidth]{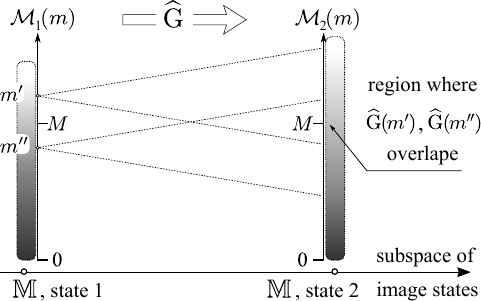}
	\end{center}
	\caption{Illustration of image transformation $\TrG$ related, e.g., to mental operations with images in copying, retaining to and retrieving from memory.}
	\label{Fig:Transf}
\end{figure}

Such mental operations may be treated as the movement of the $\mathcal{M}(n)$-cloud between different states of the image $\mathbb{M}$,  i.e., between different points of subspace composed of the possible image states.  Figure~\ref{Fig:Transf} illustrates this operation. In general, this operation can be written in the form 
\begin{equation}\label{A1:11}
	\partial \mathcal{M}_2\big(m/M,\Delta_2\big) = \TrG\Big[\partial \mathcal{M}_1\big(m/M,\Delta_1\big)\Big]\,.
\end{equation}
Accepting that  
\begin{itemize}[topsep=0.15\baselineskip, itemsep=0.1\baselineskip,  parsep=0.15\baselineskip]
	\item[-] the individual blur $\TrG(m)$ of points $m\in\mathbb{M}$ caused by the given transformation and the original $\mathcal{M}_1(m)$-cloud as a whole are governed by the same mechanism and   
	\item[-] the superposition principle for the overlapping of images $\TrG(m')$ and $\TrG(m'')$ in the formation of the $\mathcal{M}_2(m)$-cloud holds
\end{itemize}
transformation~\eqref{A1:11} can be represented in the linear form\footnote{The detailed construction of this transformation deserves an individual analysis to be published somewhere else.}
\begin{equation}
\label{A1:11LF}
\partial \mathcal{M}_2\big(m/M,\Delta_2\big)  = \int\limits_0^\infty dm'\, G(m,m'|\Delta_G)\cdot\partial \mathcal{M}_1\big(m'/M,\Delta_1\big)\,,
\end{equation}
where $\Delta_G$ is a given parameter of this transformation and
\begin{equation*}
G(m,m'|\Delta_G) \propto \int\limits_0^\infty dM'\,\partial\mathcal M\big(m/M',\tfrac{1}{\sqrt2}\Delta_G\big)\cdot \partial \mathcal M\big(m'/M',\tfrac1{\sqrt2} \Delta_G\big)\,.
\end{equation*}
Whence it follows that the transformation $\TrG$ can be reduced to the renormalization of the relative thickness of image boundary
\begin{equation}\label{A1:11Delta}
\TrG:\  \mathcal{M}_1(m) \mapsto \mathcal{M}_2(m) \Rightarrow \Delta_2 = \sqrt{\Delta_1^2+\Delta_G^2}\,.
\end{equation}
where the parameter $\Delta_G$ admits the interpretation as an additional blur caused by memory-based mental operations and attributed to the memory on its own.  

For simplicity, the memory-caused blur $\Delta_G\gg \Delta_e$ will be regarded as a fixed parameter for all the mental operations with just emerged images. Therefore all the images except for just emerged ones will be characterized by a relative boundary thickness $\Delta_m/\sqrt2\gg \Delta_e$. In this case the memory-based comparison of images $\mathbb{M}_1$ and $\mathbb{M}_2$ is specified by the probability
\begin{equation}\label{A1:12memory}
\bra{\mathbb{M}_1}\ket{\mathbb{M}_2} = \exp\big\{-\mathfrak{B}(M_1/M_2,\Delta_m) \big\}
\end{equation}
of their equality in magnitude.  Here the proximity potential $\mathfrak{B}(M_1/M_2,\Delta_m)$ is again specified by Exp.~\eqref{A1:2}. 

The similarity of Exps.~\eqref{A1:12} and \eqref{A1:12memory} within the replacement $\Delta_e\leftrightarrow\Delta_m$ enables me to employ the same formalism for describing the image comparison when the physical stimuli are active simultaneously or it is based on memory. The case of comparing images different in the relative thickness of their boundaries, $\Delta_1\neq\Delta_2$, requires an individual consideration. The comparison of images when one of them is caused by the currently active stimuli and the other is retrieved from memory exemplifies the latter case.

\subsection{Image qualitative comparison}

The relations ``$\prec$,'' ``$\succ$,'' and ``$\simeq$'' between two images $\mathbb{M}_1$, $\mathbb{M}_2$ are specified as follows. Using the proximity potential $\mathfrak{B}(M_1/M_2, \Delta_{e,m})$ the equality of two images $\mathbb{M}_1$ and  $\mathbb{M}_2$ in magnitude, $\mathbb{M}_1\simeq \mathbb{M}_2$, is accepted with the probability
\begin{equation}\label{Mod:3}
P_{12}^{\text{eq}} = \exp\big\{-\mathfrak{B}(M_1/M_2, \Delta_{e,m}) \big\}.
\end{equation}
When the equality of these images is rejected, the probability of this event is $1 - P_{12}^{\text{eq}}$,  the right relation between the magnitudes, $M_1 < M_2$ or $M_1 > M_2$ is selected. In other words, for a given pair of images $\mathbb{M}_1$ and $\mathbb{M}_2$
\begin{align}
\nonumber
	\mathbb{M}_1 &\simeq \mathbb{M}_2&  &\text{with the probability $P_{12}^{\text{eq}}$,}\\
\label{Mod:5}
\mathbb{M}_1 &\prec \mathbb{M}_2&   &\text{with the probability $1-P_{12}^{\text{eq}}$ for $M_1 < M_2$,}	\\
\nonumber	
	\mathbb{M}_1 &\succ \mathbb{M}_2&   &\text{with the probability $1-P_{12}^{\text{eq}}$ for $M_1 > M_2$,}
\end{align}
where the probability $P_{12}^{\text{eq}}$ is given by the proximity potential, Exp.~\eqref{Mod:3}.

\subsection{Integer fusion of images}

\begin{figure}
	\begin{center}
		\includegraphics[width=0.9\columnwidth]{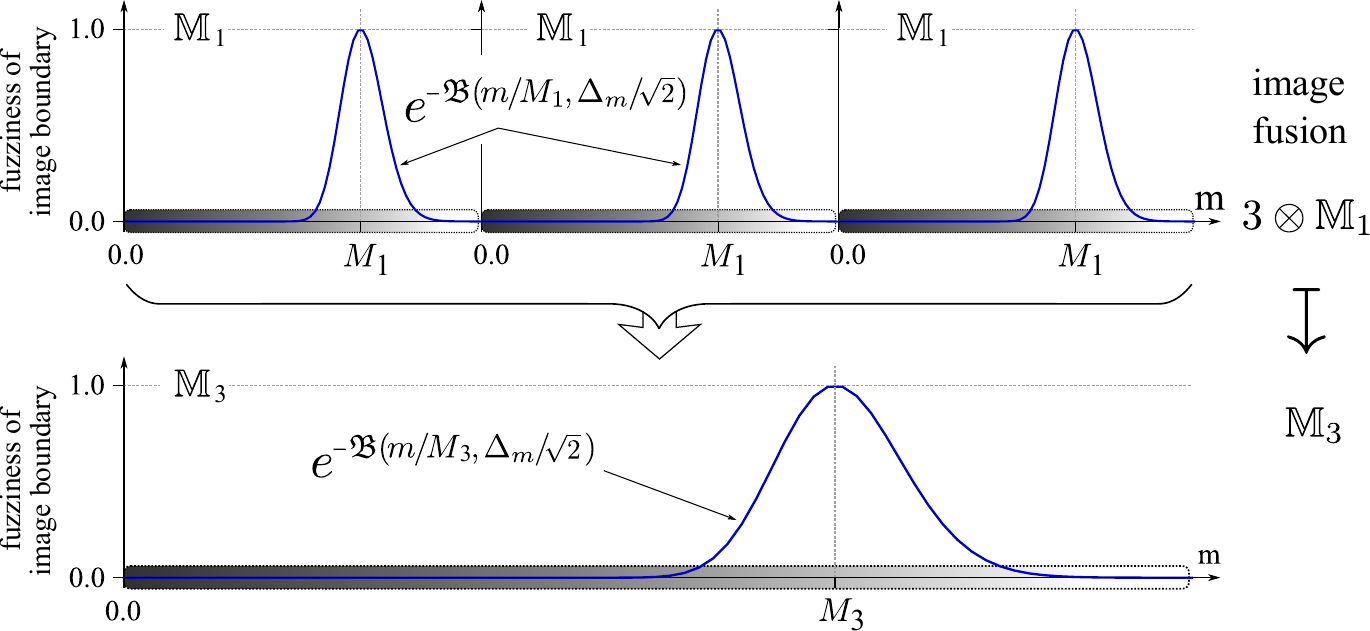}
	\end{center}
	\caption{Illustration of image fusion: three similar images $\{\mathbb{M}_1\}$ (three copies of one image) are merged into the image $\mathbb{M}_3$. The image fusion changes the scales but not the form of images.}
	\label{Fig:Fusion}
\end{figure}

Rejecting the feasibility of quantifying image magnitudes relative to one another in the standard way,  I loose the possibility of introducing the addition and subtraction operations with images. There is only one exception, it is the combination (fusion) of several images of the same magnitude (or seeming equal in magnitude) into one image made within one step (Fig.~\ref{Fig:Fusion}):
\begin{equation}\label{Mod:4}
     n\otimes \mathbb{M}_1   \mapsto \mathbb{M}_n\quad\text{implying $nM_1 = M_n$ in the sense of relations \eqref{Mod:5},}
\end{equation}
where $n =1,\, 2,\, 3,\,\ldots,\, n_c$ is some integer. This procedure, however, cannot not be introduced for any integer, it is justified only for integers less than or equal to some critical value $n_c$. As illustrated in Fig.~\ref{Fig:Fusion}, image fusion changes the scales but not the form of images, in particular, the value of $\Delta_m$. Because the given fusion is based on mental operations---mental copy of images and then conversion of their combination into one image---their boundary fuzziness is set equal to $\Delta_m/\sqrt2$. So the conscious control over the image fusion is feasible only when the uncertainty $\delta M_n$ of the result is less than the individual magnitude $M_1$ of the constituent components. In other words, we should recognize the difference of fusing $n$ and $n+1$ images for this operation to be feasible. Whence it follows that the maximal number $n_c$ of similar images that can fused is related to $\Delta_m$ as
\begin{equation}\label{Mod:6}
n_c\, \Delta_m \sim 1\,.
\end{equation}
According to the results to be discussed further and turning to our causal experience the value of $n_c$ can be estimated as $n_c\sim 4$--6.   

The constructed expressions and the found estimates of the model parameters enable me to analyze particular phenomena and compare numerically the results of the proposed model with experimental data outlined in Sec.~\ref{Sec:universality}.  

\section{Qualitative bisection and its generalization to ternary choice}\label{QBTerCh}

Let us analyze the bisection problem and related issues employing the developed formalism for image comparison.
Within the classical \textit{bisection task} subjects are asked to classify a trial stimulus  with respect to its proximity to one of two given stimuli from the same perceptual continuum. This task matches the \textit{bisection problem} concerning the image $\langle \mathbb{M}\rangle$ whose magnitude $\langle M\rangle$ admits the interpretation as the mean value of the magnitudes $M_1$ and $M_2$  of other two images $\mathbb{M}_1$ and $\mathbb{M}_2$.  Below in this section I will assume that $\mathbb{M}_1 \prec \mathbb{M}_2$, because for the bisection problem to have some sense the qualitative difference between the two images has to be well recognizable.

The particular expression for $\langle \mathbb{M}\rangle$ depends on several factors, with two of them playing crucial roles. First, it is whether a subject evaluates the difference between the magnitudes $M_1$ and a trial $m$, on one side, and $m$ and $M_2$, on the other side, qualitatively, i.e., without any attempt to compare these differences quantitatively. For this reason I use the term \textit{qualitative bisection} to refer to such image (signal) comparison. As clear from the results of Sec.~\ref{sec:CCIP}, the \textit{quantitative} comparison can become rather problematic if the ratio $M_2/M_1\lesssim 2$. 

The second crucial factor is whether a subject is able to quantify the magnitude $M_2$ in units of $M_1$. If not so, the contribution of the magnitude $\mathbb{M}_1$ into the value of $\langle M\rangle$ is just ignorable and the subject can quantitatively construct the image $\langle \mathbb M\rangle$  with the magnitude $\langle M\rangle \approx M_2/2$. Under such conditions the magnitude  $\langle M\rangle$ may be classified as the \textit{arithmetic mean}.
According to Sec.~\ref{Sec:universality} it is the case when the ratio of their magnitudes exceeds $n_c$, i.e., $M_2\gtrsim n_c M_1$. 

Within the qualitative bisection problem the proximity potential $\mathfrak{B}$ is used to quantify the nearness of two images, which leads to the definition of the mean value $\langle  M\rangle$ as a magnitude meeting the condition 
\begin{equation}\label{QB:1}
       \mathfrak{B}\bigg(\frac{m}{M_1},\Delta_m\bigg)\bigg|_{m= \langle \mathbb M\rangle} =   \mathfrak{B}\bigg(\frac{M_2}{m},\Delta_m\bigg)\bigg|_{m= \langle \mathbb M\rangle}\,,
\end{equation}
whence it directly follows that
\begin{equation}\label{QB:2}
\langle \mathbb M\rangle =  \langle \mathbb M\rangle_g \stackrel{\text{def}}{{}={}}  \sqrt{M_1\,M_2}\,.
\end{equation}
Thereby the qualitative bisection is characterized by the geometric mean. 

It is worthy of noting that arguments giving rise to equality~\eqref{QB:2} are not new, at least, they are met in publications by Luce~\cite{luceMathBio1963.ch5,Luce1961}. What concerns available experimental data for stimulus bisections, let us turn to psychophysics of time perception because the magnitude of perceived duration and the real duration of various stimuli are rather similar \cite[][for a review]{Wearden2016psychology}. A meta-analysis of classification data \cite{KOPEC2010262} reports that the bisection point lies near the geometric mean for ratios of referent duration about 2 or less and lies near the arithmetic mean for ratios of 4 or greater. The noted values of referent ratio fit well the value of $n_c$ estimated above as $n_c\sim4$--6. For a detailed review of the modern state of the art in this field a reader may be referred to Ref.~\cite{fnint.2015.00044}.

The qualitative bisection problem admits a certain generalization when a given magnitude $m$ has to be classified according to its proximity to one of two magnitudes $M_1$ and $M_2$ ($M_1 < m < M_2$), whereas
\begin{itemize}[topsep=0.15\baselineskip, itemsep=0.1\baselineskip,  parsep=0.15\baselineskip]
	\item[-] the difference between the trial magnitude $m$ and both of the classification magnitudes $M_1$ and $M_2$ is clearly recognized, i.e.,
	\begin{equation*}
	   \mathfrak{B}_i = \mathfrak{B}(m/M_i,\Delta_m) \gg 1\quad (i=1,2),
	\end{equation*}
	\item[-] there is no other initially given magnitude between $M_1$ and $M_2$ that can be used as a classification value.  
\end{itemize}
This situation can be met in quantitative comparison of images, so let us analyze it here in more details.

Referring to Exp.~\eqref{Mod:3} as some analogy to the analyzed problem the \textit{binary} choice between $M_1$ and $M_2$ is described by the probability of choosing $M_i$  ($i=1,2$) equal to
\begin{equation}\label{QB:3}
	P_i = \frac1Z e^{-\mathfrak{B}_i}
\end{equation}
where the normalizing coefficient $Z$ is specified by the expression 
\begin{equation*}
Z = e^{-\mathfrak{B}_1} + e^{-\mathfrak{B}_2}
\end{equation*}
because only the two options are available for selection. However, although no other options are given initially there is, nevertheless, another universal choice tacitly present. It is to reject both the options $M_1$ and $M_2$ when $\mathfrak{B}_1,\mathfrak{B}_2\gg 1$. If, in any case, the choice of some classification magnitude has to be made, the selection of ``something'' between $M_1$ and $M_2$ can be used as the implementation of this choice. It introduces a new classification option---some mean value $\langle \mathbb M\rangle_{12}$---in consideration. Leaping ahead, I note that this value may be identified with the geometric mean~\eqref{QB:2} but the proximity potential $\mathfrak{B}(m/\sqrt{M_1M_2},\Delta)$ formally constructed for $m$ and  $\langle M\rangle_g$ cannot be used here to quantify the nearness of the trial magnitude $m$ to the value $\langle M\rangle_g$. The matter is that according to the initial conditions there can be no image  created in some way whose magnitude is between the values $M_1$ and $M_2$. Thereby no specific magnitude can be attributed to the ``something'' between $M_1$ and $M_2$. So in describing the choice of the mean value $\langle M\rangle_g$ as the alternative class of magnitude classification I am allowed to deal only with the quantities $\mathfrak{B}_1(m)$ and $\mathfrak{B}_2(m)$ considered, maybe, for all the possible values of trial magnitude, $m\in[M_1,M_2]$. 

Let us introduce the value $\mathfrak{B}_g$ called, by analogy, the proximity potential to describe the choice of  $\langle M\rangle_g$ in the same way as it is done for the magnitudes $M_1$ and $M_2$. Namely, let the probability of choosing $\langle M\rangle_g$ be
\begin{equation}\label{QB:4}
P_g = \frac1Z e^{-\mathfrak{B}_g}
\end{equation}
where now by virtue of Exps.~\eqref{QB:3} and \eqref{QB:4} the normalization coefficient  $Z$ is specified by the expression 
\begin{equation}\label{QB:5}
Z = e^{-\mathfrak{B}_1} + e^{-\mathfrak{B}_2} + e^{-\mathfrak{B}_g}.
\end{equation}
In order to construct a plausible form of the dependence $\mathfrak{B}_g[\mathfrak{B}_1,\mathfrak{B}_2]$ I note the following. For the general reasons when, e.g., $\mathfrak{B}_1\ll 1$, the use of the mean value class is not necessary, which can be taken into account expecting that $\mathfrak{B}_g\gg 1$ in this case.  Otherwise, when $\mathfrak{B}_1,\mathfrak{B}_2\gg 1$ the choice of $\langle M\rangle_g$ may be dominant, which implies the inequality $\mathfrak{B}_m\ll \mathfrak{B}_1,\mathfrak{B}_2$ to hold. Finally, when the trial magnitude $m\sim \langle M\rangle_g$ and $\mathfrak{B}_1(m) \sim \mathfrak{B}_2(m)\sim 1$, selecting all the three magnitude classes must be equiprobable, i.e.,  $\mathfrak{B}_g\sim 1$ also. If for such trial magnitudes $m$ the values $\mathfrak{B}_1(m)  \sim \mathfrak{B}_2(m)\gg 1$ the proximity potential $\mathfrak{B}_g(m)$ (\textit{i}) has to take small values, $\mathfrak{B}_g\ll1$, and (\textit{ii}) should be the minimum achieved by the function $\mathfrak{B}_g(m)$ for $m\in[M_1.M_2]$.   

\begin{figure}
	\begin{center}
		\includegraphics[width=0.85\columnwidth]{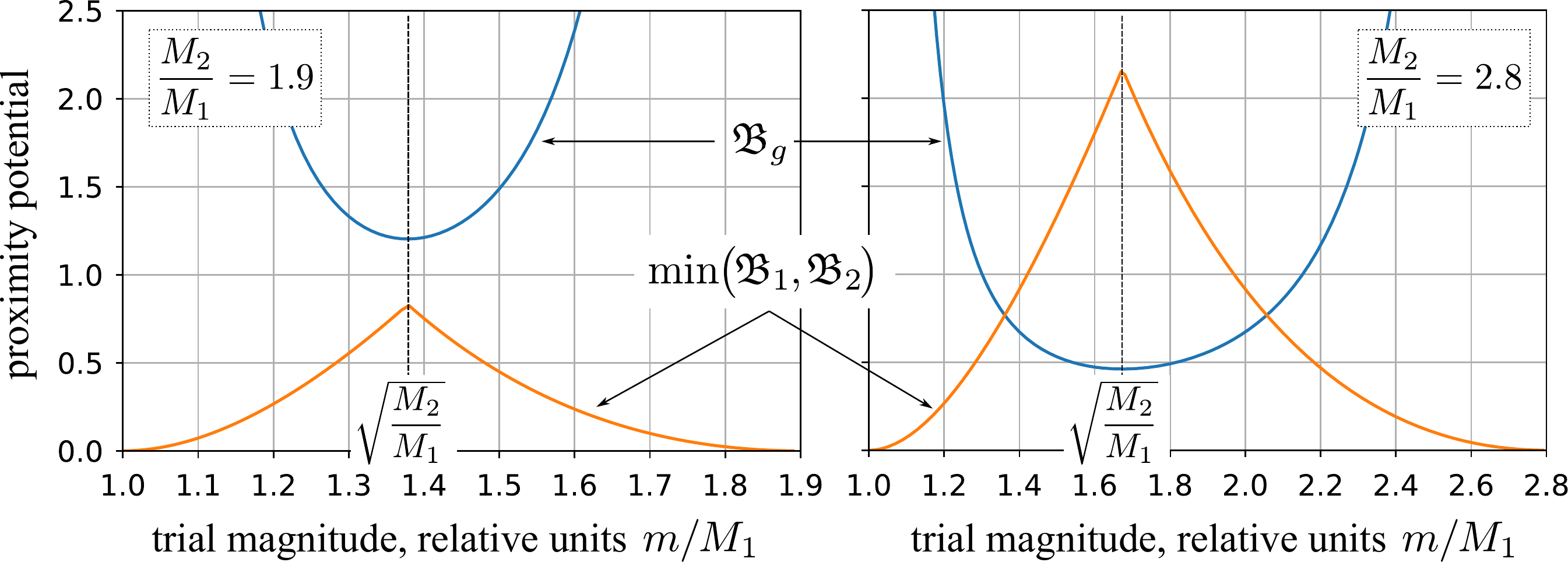}
	\end{center}
	\caption{Qualitative bisection problem generalized to the ternary choice between two initially given magnitudes $M_1$, $M_2$  and their mean value $\langle M\rangle_g$ as the rejection of both the options  $M_1$, $M_2$. In plotting the proximity potentials $\mathfrak{B}_1(m)$, $\mathfrak{B}_2(m)$, and $\mathfrak{B}_g(m)$ based on Exps.~\eqref{A1:2}, \eqref{Mod:3}, \eqref{QB:6} the parameter $1/(2\Delta_m^2) = 8$ and $\mathfrak{B}_0= 1$ were used, which corresponds to $(M_2/M_1)_c\approx 2.02$}
	\label{Fig:Bisection}
\end{figure}

The following simple ansatz 
\begin{equation}\label{QB:6}
  \mathfrak{B}_g(m) = \frac{\mathfrak{B}_0^2}2\bigg[\frac1{\mathfrak{B}_1(m)} + \frac1{\mathfrak{B}_2(m)}\bigg]
\end{equation}
meets all these requirements. Here I have introduced the additional parameter $\mathfrak{B}_0$ playing the role of fuzzy threshold evaluating the proximity potentials $\mathfrak{B}_1(m)$ and  $\mathfrak{B}_2(m)$  when the contribution of the mean value choice becomes remarkable. In the preceding argumentation this threshold was tacitly set equal to unity because of exponential decrease of probability~\eqref{QB:3} for $\mathfrak{B}_i\gg 1$. However, there could be other reasons for turning to this option.

Figure~\ref{Fig:Bisection} illustrates the transition from practically the binary choice to the situation where the contribution of the mean-value option is crucial. As follows from Exp.~\eqref{QB:6} there is a critical ratio $r_c = (M_2/M_1)_c$ of magnitudes specified by the expression
\begin{equation}
    r_c = \frac14\Big(\Delta_m\sqrt{\mathfrak{B}_0} + \sqrt{2+\Delta_m^2\mathfrak{B}_0}\,\Big)^4\,,
\end{equation}
such that for $M_2/M_1 < r_c$ the choice between the magnitudes $M_1$ and $M_2$ does not imply a substantial contribution of the mean-value option. If $M_2/M_1 > r_c$, the choice of any of the classification magnitudes $M_1$ and $M_2$  at least for  $m\sim \langle M\rangle_g$ is not relevant and the possibility of referring to the mean value as a classification option becomes essential.

When the choice of alternatives to the initially given options is not feasible the developed model can be employed for describing the rejection of the choice at all. We will meet this situation further when two magnitudes $M_1$ and $M_2$ that should be compared quantitatively become incommensurable.

\section{Direct quantitative comparison of images: Bounded rational arithmetic}\label{Sec:QCI}

Quantitative comparison of two images $\mathbb{M}_1$, $\mathbb{M}_2$ with respect to their magnitudes $M_1$ and $M_2$ requires some mental operation giving the ratio $M_1/M_2$ as a number. The possibility of implementing such operations unconsciously, at least, concerning their main steps seems to be doubtful  because mental images are certain objects of mental space rather than abstract numbers. Contrarily, relations~\eqref{Mod:P3a} and integer fusion~\eqref{Mod:P3b} deal with images as whole objects, so the introduction of these operations at the general level of mental space description is acceptable. 

In the present section I demonstrate that the desired quantitative comparison of images admits a conscious implementation  based on  the two operations using some common unit---an image $\mathds1$ commensurable  with the analyzed images. 

Let us consider two images $\mathbb{M}_1$ and $\mathbb{M}_2$ of magnitudes $M_1$ and $M_2$ representing stimuli of the same perceptual continuum and a similar trial image $\mathds1$ of  amplitude $\ell$ to be used as a common unit for quantifying the two ones. Independently of whether the analyzed images are caused by currently active stimuli or, at least one of them is retrieved from memory, the image fusion is a pure mental operation. Therefore for describing the relationship between these images the proximity potential $\mathfrak{B}(\ldots,\Delta_m)$ will be employed in further analysis.

The \textit{direct quantitative comparison} of the two images  $\mathbb{M}_1$ and $\mathbb{M}_2$ via the trial unit $\mathds{1}$ is based on the attempt:
\begin{itemize}[topsep=0.15\baselineskip, itemsep=0.1\baselineskip,  parsep=0.15\baselineskip]
	\item[-] to fuse $n_1$ copies of $\mathds{1}$ and $n_2$ copies of $\mathds{1}$,
	\item[-] to compare pairwise the results and the images  $\mathbb{M}_1$ and $\mathbb{M}_2$ qualitatively with respect to their equality in magnitude and then
	\item[-] to find the optimal configuration $\{M_1,M_2, \ell, n_1,n_2\}$ which determines the \textit{quantitative estimation} of  magnitude ratio
	\begin{equation*}
	\bigg(\frac{M_2}{M_1}\bigg)_e \stackrel{\text{def}}{{}={}} \frac{n_2}{n_1}\bigg|_{\text{opt}}.
	\end{equation*}
\end{itemize} 
The following diagram 
\begin{equation}\label{Mod:9}
\begin{CD}
	n_1\otimes\mathds1\simeq\mathbb{M}_1    @<<< \mathds1 @>>>  n_2\otimes\mathds1\simeq\mathbb{M}_2\\
	@.   @VVV       @.\\
	M_1   @>>>  \ket{}\cdot\, \dfrac{n_2}{n_1}  @>>>  M_2
\end{CD}  
\end{equation}
illustrates this procedure. The appropriate choice of trial units ensures the integers $n_1$ and $n_2$ to be relatively prime. Besides they must meet the inequality
\begin{equation}\label{Mod:9add}
    n_1,\, n_2 \leq n_c\,.
\end{equation}
required for the integer fusion of images to be well defined. The latter inequality is the reason to call the given procedure the magnitude quantification within \textit{bounded rational arithmetic}.

By virtue of \eqref{A1:2} and \eqref{Mod:3} the probability $\mathcal{P}(n_1,n_2,\ell)$ of considering the images $\mathbb{M}_1$ and $\mathbb{M}_2$ equivalent in magnitude based on the operations shown in diagram~\eqref{Mod:9} is given by the expression
\begin{equation}\label{Mod:10}
    \mathcal{P}(n_1,n_2,\ell) = \exp\big\{   
     - \mathcal{B}_\ell(n_1,n_2)
    \big\},
\end{equation}
where the $\ell$-dependent proximity potential $\mathcal{B}_\ell$ is composed of two terms
\begin{equation}\label{Mod:11}
\mathcal{B}_\ell(n_1,n_2) =\frac1{2\Delta_m^2} \left[\sqrt{\frac{M_1}{n_1\ell}} - \sqrt{\frac{n_1\ell}{M_1}}\right]^2 
+ \frac1{2\Delta_m^2} \left[\sqrt{\frac{M_2}{n_2\ell}} - \sqrt{\frac{n_2\ell}{M_2}}\right]^2 .
\end{equation}
Due to estimate~\eqref{Mod:6} and condition~\eqref{Mod:9add} acceptable variations of $\ell$ near its optimal value
\begin{equation*}
\ell_\text{opt} = \sqrt{\frac{M_1}{n_1}\,\frac{M_2}{n_2}}
\end{equation*}
matching the minimum of $\mathcal{B}_\ell$ are less than $\ell_\text{opt}/n_c$, which allows us to use the replacement  $n_i\ell\to n_i\ell_\text{opt}$ ($i=1,2$) in Exp.~\eqref{Mod:11}. In this way we get the probability $\mathfrak{P}$ of quantifying the image amplitudes $M_1$ and $M_2$ in terms of the identity $M_1/M_2 \simeq n_1/n_2$ via the equality
\begin{equation}\label{Mod:12}
\mathfrak{P}\Big(\frac{n_1}{n_2}\,\Big|\, \frac{M_1}{M_2}, \Delta_m\Big) \stackrel{\text{def}}{=} \frac1{Z}\exp\big\{-\mathfrak{B}_n
(\rho,\Delta_m)\big\} ,
\end{equation}
where the proximity potential $\mathfrak{B}_n$ of the configuration $\{n_1,n_2\}$ is specified by the expression
\begin{align}
\nonumber
   \mathfrak{B}_n(\rho,\Delta_m)& = \frac1{\Delta_m^2}\left[\rho^{1/4}-\frac1{\rho^{1/4}}\right]^2 \quad \text{with}\quad \rho = \frac{M_1}{M_2}\cdot\frac{n_2}{n_1} \\
\label{Mod:13}   
   {}& \approx \frac1{4\Delta_m^2}\left[\sqrt{\rho}-\frac1{\sqrt\rho}\right]^2 \quad \text{for}\quad \Delta_m \ll 1
\end{align}
and the normalization factor $Z$ is determined by the collection of possible configurations $\{n_1,n_2\}\in \mathcal{C}$ of relatively prime integers:
\begin{equation}\label{Mod:13CC}
Z = \sum_{\{n_1,n_2\}\in \mathcal{C}}\exp\big\{-\mathfrak{B}_n(\rho_n,\Delta_m)\big\}\, .
\end{equation}
The collection $\mathcal{C}$ of possible configurations is bounded by condition~\eqref{Mod:9add} imposed on the integers $n_1$ and $n_2$ such that their maximal ratio is $n_c$. Therefore  in the given way only magnitudes $M_1$ and $M_2$ obeying the condition
\begin{equation*}
\max\left( \frac{M_2}{M_1}\right) \lesssim n_c
\end{equation*}
can be quantitatively compared.

This interval of magnitudes admitting the quantitative comparison can be expanded within the same formalism. In the presented approach several copies of a trial unit $\mathds{1}$ were fused for qualitative comparison with the images $\mathbb{M}_1$ and $\mathbb{M}_2$. So its magnitude $\ell$ cannot exceed the minimal value of the magnitudes $M_1$ and $M_2$, let it be $M_1$. However this comparison may be also implemented via merging several copies of $\mathbb{M}_1$ for qualitative comparison with a trial unit $\mathds{1}$.\footnote{It should be noted that the case when both the images $\mathbb{M}_1$ and $\mathbb{M}_1$ are merged for pairwise qualitative comparison with a trial unit $\mathds{1}$ is reduced to the initially analyzed situation  within the replacement $n_1\leftrightarrow n_2$.}
The given situation is described by the same formalism within the replacement
\begin{equation}\label{Mod:13mod}
\frac{M_1}{n_1}\rightarrow M_1 n_1 \,.
\end{equation}
Thereby, the direct quantitative comparison of two images $\mathbb{M}_1$ and $\mathbb{M}_2$ is possible when
\begin{equation}\label{Mod:Commensur}
\max\left( \frac{M_2}{M_1}\right) \lesssim n_c^2\,.
\end{equation}
Here I have used the symbol $\lesssim$ rather than $\leq$ to emphasize that the boundary of this interval is blurred  in some way. 

The images meeting condition~\eqref{Mod:Commensur} will be called \textit{directly commensurable} or just \textit{commensurable} for short, otherwise they are \textit{incommensurable}. The value $n_c^2$ will be also referred to as the \textit{capacity} of direct quantitative comparison of images or just the \textit{commensurability capacity}.

The following diagram generalizing \eqref{Mod:9} illustrates the direct quantitative comparison of images $\mathbb{M}_1$ and $\mathbb{M}_2$ and the possible choice of trial configurations   
\begin{equation}\label{Mod:9final}
\begin{CD}
\begin{aligned} a)\ n_1 &\otimes\mathds1\simeq\mathbb{M}_1 \\[-0.25\baselineskip]
              b)\ n_1 & \otimes\mathbb{M}_1 \simeq\mathds1
\end{aligned}
@<<< \mathds1 @>>>  n_2\otimes\mathds1\simeq\mathbb{M}_2\\
@.   @VV 
\ 1 \leq n_1 \leq n_2 \lesssim n_c\quad
	 V@.\\
M_1   @>>>  \ket{}\times \begin{aligned}a)&\quad\dfrac{n_2}{n_1}\\b) &\quad n_1\cdot n_2 \end{aligned}  @>>>  M_2
\end{CD}  
\end{equation}

Summarizing these results, the \textit{direct quantitative comparison} of two images $\mathbb{M}_1$, $\mathbb{M}_2$ with amplitudes $M_1$, $M_2$ admits the following probabilistic description: 
\begin{itemize}[topsep=0.15\baselineskip, itemsep=0.1\baselineskip,  parsep=0.15\baselineskip]
	\item The possible pairs $\{n_1,n_2\}$ of integers from the interval $[1, n_c]$ make up the collection of quantities 
	$\{r_n \}  = \{n_2/n_1\}\bigcup\{n_1\cdot n_2\}$ which may be used for the quantitative comparison of the amplitudes $M_1$, $M_2$.
	
	\item The fitness of a given quantity $r_n$ for estimating the magnitude ratio $M_2/M_1$ is determined by the proximity potential $\mathfrak{B}_n$ described by Exp.~\eqref{Mod:13} or its modification within replacement~\eqref{Mod:13mod}. 
	The probability $\mathfrak{P}_n$ of selecting the given quantity $r_n$ is specified by $\mathfrak{B}_n$ according to Exp.~\eqref{Mod:12}. 
	
	\item If the most probable quantity $r^{\text{opt}}_n(M_2/M_1)$ is characterized by a large value of its proximity potential $\mathfrak{B}_n\gtrsim\mathfrak{B}_0$ then the set $\{r_n\}$ has to be extended to include the mean-value option as wall as the possibility of rejecting the feasibility of the magnitude quantitative comparison at all.   
\end{itemize}
Leaping ahead, I note that this probabilistic model for image comparison can also describe the hysteresis in quantifying the magnitude relationship after a minor modification allowing for the decision inertia \cite[e.g.,][]{Akaishi2014195,carlos2016inertia}. 

\section{Commensurability capacity and inner psychophysics}\label{sec:CCIP}

Expression~\eqref{Mod:Commensur} for the capacity of direct quantitative comparison of images enables me, first, to justify Teghtsoonian's estimate of the dynamic range $R_M$ of sensory magnitude common for many perceptual continua, $\log R_M\sim 1.5$ (Sec.~\ref{Sec:universality}, Proposition~\ref{Th:2}). According to the accepted definition, the dynamic range of sensory magnitudes is the span from the lowest to highest magnitudes of a given perceptual continuum admitting a continuous grading. In the terms of developed model the dynamic range $R_M$ is just the maximal ratio of magnitudes admitting direct quantitative comparison, so
\begin{equation}\label{Res:1}
R_M \sim n_c^2\,.
\end{equation}  
Setting here $R_M \approx 10^{1.5}$ I formally get $n_c\approx 5.6$ which belongs to the interval $n_c\sim 4$--6 of values expected based on our casual experience.

Second, the value $n_c^2$ estimates the minimal magnitude $M_{\text{th},m}$ of images that can be \textit{quantitatively} compared with a given image $\mathbb{M}$ of magnitude $M$, i.e., 
\[
M_{\text{th},m}\sim \frac1{n_c^2} M\,.
\]
Within the developed model Ekman's law~\eqref{Ekmanlaw} deals with the corresponding threshold $M_{\text{th},e}=\kappa_e M$ describing the just-noticeable difference of magnitudes provided the stimuli causing the emergence of these images are currently active.  The inequality $M_{\text{th},e} \gg M_{\text{th},m}$ cannot hold because for a difference between two magnitudes to admit quantitative evaluation it has to be well recognized.
The opposite inequality $M_{\text{th},e} \ll M_{\text{th},m}$ seems to be in direct contradiction to the self-consistency of human sensory system. Indeed the physiological necessity of magnitude discrimination admitting qualitative implementation but not quantitative evaluation is unclear.   
So it is quite reasonable  to hypothesize that the two thresholds should be similar, i.e., the equality 
\begin{equation}\label{Res:2}
\kappa_e \sim \frac1{n_c^2}
\end{equation}  
has to hold. Using Teghtsoonian's estimate of the Ekman fraction $\kappa_e\approx 0.03$ (Sec.~\ref{Sec:universality}, Proposition~\ref{Th:1})  Exp.~\eqref{Res:2} immediately gives $n_c\sim 5.8$. The latter again is rather close to the values noted above.  Whence it also follows that the parameters $\Delta_e$ and $\Delta_m$ entering the proximity potential for images when their stimuli are currently active and when their comparison is based on memory are related as $\Delta_e\sim \Delta_m^2$. 

\section{Direct quantitative comparison of images: Intrarange properties}

In the present section I discuss the regression and range effects as well as the hysteresis in magnitude quantitative comparison within the range of direct commensurability.\footnote{The related experimental data were described  in Sec.~\ref{Sec:universality}, Propositions~\ref{Th:4}, 5.} To be determined, let us consider two images $\mathbb{M}_1$ and $\mathbb{M}_2$ of magnitudes $M_1$ and $M_2$ when $M_1 <M_2$ and the image $\mathbb{M}_1$ is regarded as the standard object whereas the image $\mathbb{M}_2$ it trial one. 

Following the model for direct quantitative comparison of images developed in Sec.~\ref{Sec:QCI}, the collection of quantities $\mathcal{C}=\{r_n\}$ that may be used for quantifying the magnitude ratio $M_2/M_1$ is constructed and the corresponding value $\mathfrak{B}_n =\mathfrak{B}(M_2/M_1,r_n)$ of the proximity potential is calculated for each $r_n\in\mathcal{C}$.  

\begin{figure}
	\begin{center}
		\includegraphics[width=\columnwidth]{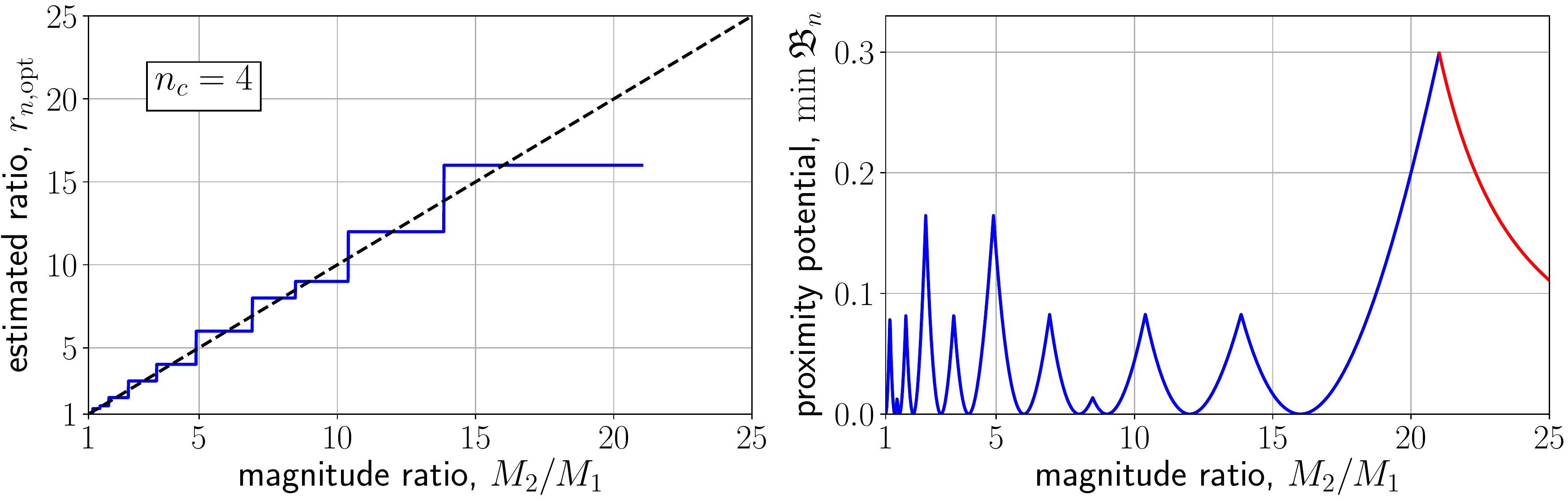}
	\end{center}
	\caption{The choice of the evaluated ratio $r_{n,\text{opt}}$ for magnitudes $M_2$ and $M_1$ determined by the minimum of proximity potential $\min\mathfrak{B}_n$ (left panel) and this potential minimum vs the ratio $M_2/M_1$ (right panel). Red line represents the potential attributed to rejecting the feasibility of image comparison, in the presented data the parameters $n_c=4$ and $\mathfrak{B}_0 = 0.3$ were used (Sec.~\ref{QBTerCh}).}
	\label{Fig:Ideal}
\end{figure}

\subsection{Perfect rationality approximation}

If the choice of quantities $\{r_n\}$ were perfectly rational, then the optimal value $r_{n,\text{opt}}$ matching the minimum of the proximity potential, $\min\mathfrak{B}_n$, within the collection $\{\mathfrak{B}_n\}$ would be selected. Figure~\ref{Fig:Ideal}\,(left panel) depicts the corresponding step-wise variations in the quantity $r_{n,\text{opt}}$  within the region $M_2/M_1 \lesssim n_c^2$, where the magnitudes $M_1$ and $M_2$ are commensurable (here the value $n_c = 4$ has been used).

The right panel shows the corresponding pattern of proximity potential. As seen, the variations in $\min\mathfrak{B}_n$ are about or less 0.1 inside the region $M_2/M_1 < n_c^2$. Therefore the choice of ``mean'' value as an additional class of categorization may be ignored (Sec.~\ref{QBTerCh}). However, when the magnitude ratio $M_2/M_1$ goes outside the region $[1,n_c^2]$ the value of $\min\mathfrak{B}_n$ starts to grow and finally exceeds the threshold $\mathfrak{B}_0$. In this case the feasibility of quantitative comparison of the two magnitudes has to be rejected and these magnitudes become incommensurable. In the presented analysis I have estimated the threshold $\mathfrak{B}_0$ as a quantity exceeding to some degree the characteristic variations in $\min\mathfrak{B}_n$ for internal values $M_2/M_1\in [1,n_c^2]$ and set $\mathfrak{B}_0=0.3$ for $n_c=4$. 

\subsection{Probabilistic features of magnitude quantification}

\begin{figure}
	\begin{center}
		\includegraphics[width=\columnwidth]{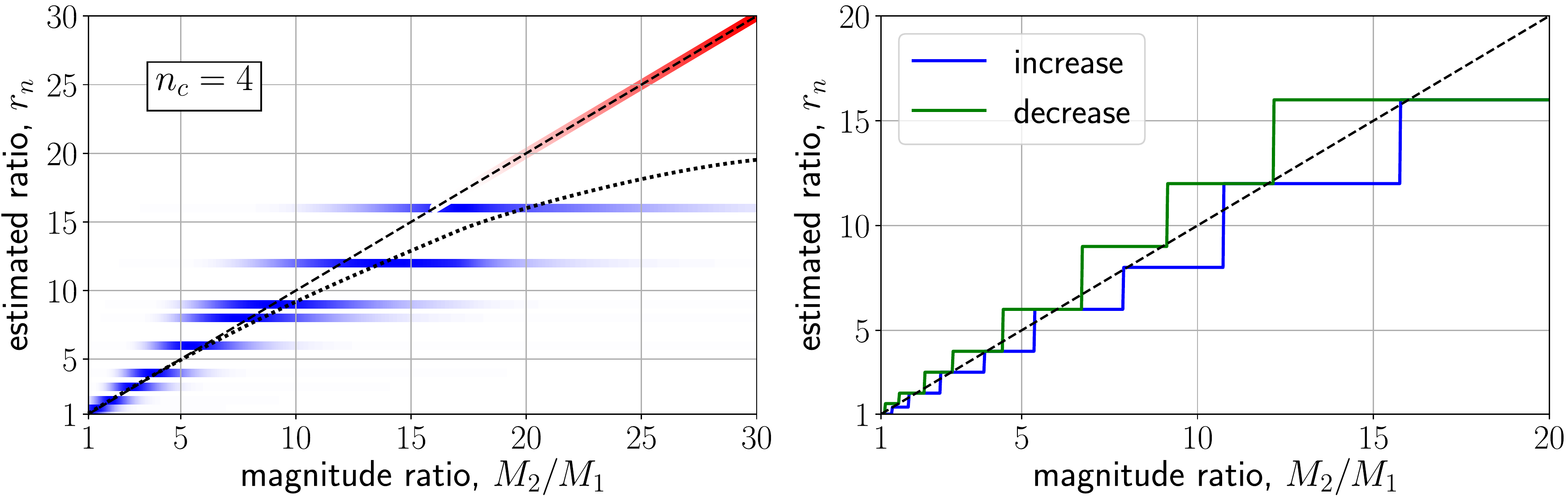}
	\end{center}
	\caption{The probabilistic pattern of $r_n$-choice in the quantitative comparison of magnitudes $M_1$ and $M_2$ (left panel) and the hysteresis effected presented within the perfect rationality approximation (right panel). The presented results were obtained for $n_c=4$ and $\mathfrak{B}_0= \mathfrak{B}_i = 0.3$.}
	\label{Fig:Probab}
\end{figure}

The probabilistic features of magnitude comparison described by the developed model are illustrated in  Fig.~\ref{Fig:Probab}. As previously, for a given ratio $M_2/M_1$ the proximity potential $\mathfrak{B}_n$ is calculated for each quantity $r_n$. However the choice of a value $r_n$ also allowing for the possibility of rejecting the set of options $\{r_n\}$ as a whole is described by the probability given by Exp.~\ref{Mod:12}. The result is shown in  Fig.~\ref{Fig:Probab}\,(left panel), where the blue strips match the quantities  $\{r_n\}$ and the saturation of their color visualizes the probability of choice according to the formula
\[
	\text{saturation}_n\propto \bigg\{P_n(M_2/M_1)\cdot\Big[\max_{M_2/M_1} P_n(M_2/M_1)\Big]^{-1}\bigg\}^2.
\]
In the same way the red strip visualizes the rejection of image comparison. 

As seen, the position of fuzzy boundary between the regions of magnitude commensurability and incommensurability can be estimated as $(M_2/M_1)_c \sim 25$. The later value may be treated as the maximal dynamic range of sensory magnitudes and fits well its estimates $10^{1.5}\sim 30$ obtained for the available experimental data (Sec.~\ref{Sec:universality}), which also justifies the used parameter $n_c = 4$. 

The obtained results allow me to state that the model for direct quantitative comparison of images developed in Sec.~\ref{Sec:QCI} is characterized by the following features.
\begin{itemize}[topsep=0.15\baselineskip, itemsep=0.1\baselineskip,  parsep=0.15\baselineskip]
	\item \textit{Fuzziness of direct commensurability}. The bounded capacity of human cognition reflected in the probabilistic mechanism of image comparison endows the region of magnitude direct commensurability---the maximal dynamic range of sensory magnitudes---with fuzzy boundaries. 
	
	\item \textit{Regression effect.} The probabilistic choice of \textit{individual} values $r_n\in \mathcal{C}$ in evaluating the quantitative relationship between magnitudes $M_1$ and $M_2$ is biased towards the underestimation of the ratio $M_2/M_1$ in the analyzed case.\footnote{This conclusion is drawn for $M_2 > M_1$. In the opposite case, $M_2 < M_1$, it converts into the overestimation of the ratio $M_2/M_1$.}  
	This underestimation seems to be caused by that the number of quantities $\{r_n^<\}$ located under the value $M_2/M_1$ exceeds the number of similar quantities $\{r_n^>\}$ above it for $M_2/M_1\sim n_c^2$.

	\item \textit{Range effect}. The bias noted above becomes more pronounced as the magnitude ratio $M_2/M_1$ increases, which is illustrated by dotted line in  Fig.~\ref{Fig:Probab}~(left panel).
	
	\item \textit{Scaling}. Because the ratio $M_2/M_1$ of magnitudes rather than their individual values determines the image comparison the absolute value of the underestimation  (overestimation) should increase linearly, at least in the first approximation, with growth of magnitudes. If the relative thickness of $\mathcal{M}(m)$-cloud boundary $1/n_c$ increases with magnitude this increase can become super-linear.  

\end{itemize}

The choice of the relevant values from the collection $\{r_n\}$ can be affected by the decision inertia---the preference for the currently selected option \cite[e.g.,][]{Akaishi2014195,carlos2016inertia}. I model this effect ascribing an additional preference potential $\mathfrak{B}_i$ to the currently selected value $r_{\text{cur}}$. Thereby the probability of selecting a value $r_n$ is represented as 
\begin{equation}\label{mod:inertia}
   \mathfrak{P}_n = \frac1Z \exp\Big[-\mathfrak{B}_n + \delta(r_n,r_{\text{cur}})\, \mathfrak{B}_i \Big]\,,
\end{equation}
where $\delta(r,r')$ is the Kronecker symbol. In this case as the test magnitude $M_2$ gradually increases or decreases the change of the previously selected estimate $r_{\text{cur}}$ requires overcoming the potential barrier $\mathfrak{B}_i$. So, on the average, the dependence $r_n^+(M_2/M_1)$ corresponding to the gradual increase in $M_2/M_1$ has to lie below a similar dependence $r_n^-(M_2/M_1)$ for the gradual decrease. 

Figure~\ref{Fig:Probab}~(right panel) justifies this statement. The shown dependences were constructed within the approximation dealing with the minimal proximity potential modified as shown by Exp.~\ref{mod:inertia}. In plotting these data I set $\mathfrak{B}_i = \mathfrak{B}_0$ assuming that the preference of keeping the previously selected value $r_\text{cur}$ and the readiness for rejecting the image comparison may be of the same mechanism. 

The given result allows me to state that the model for direct quantitative comparison of images after its modification allowing for decision inertia admits also hysteresis in magnitude quantification. Namely: 
\begin{itemize}[topsep=0.15\baselineskip, itemsep=0.1\baselineskip,  parsep=0.15\baselineskip]

	\item\textit{Hysteresis}. The effect of decision inertia endows the image comparison with memory effects, in particular, makes the dependence  $r_n(M_2/M_1)$ sensitive to the direction of magnitude gradual variations. 
  
\end{itemize}

All the features noted above in this section are due to the discreteness of values $\{r_n\}$ available for quantifying magnitude relations, which reflects the bounded capacity of human cognition.

\section{Multi-step comparison of images and the Fechner-Stevens dilemma}

\subsection{Multi-step comparison}

When the magnitudes $M_1$ and $M_2$ of two images $\mathbb{M}_1$ and $\mathbb{M}_2$ differ substantially such that, e.g., $M_2\gg n_c^2 M_1$ their direct quantitative comparison becomes impossible because there is no unit $\mathds{1}$ commensurable with both of them. In this case, however, the comparison can be implemented via several steps with intermediate units pair-wise commensurable as illustrated in the following diagram for three-step comparison:
\begin{equation}\label{FS:1}
\begin{CD}
\text{estimated:\ }@.M_1 @>r_1>r_1>  m'@>r_2>r_1r_2> m'' @>r_3>r_1r_2r_3> m_2\\
\mathbb{M}_1  @.   @|  @AA\mathds{1}' A @AA\mathds{1}'' A  @AAA \ \mathbb{M}_2   \\
\text{given:}  @.  M_1  @>r_1q_1>r_1q_1>  M' @>r_2q_2>r_1r_2\cdot q_1q_2>  M'' @>r_3q_3>r_1r_2r_3\cdot q_1q_2q_3>  M_2
\end{CD}  
\end{equation}
Here $r_1$ is the quantity from the collection $\{r_n\}$ taken consciously to quantify the relationship between the magnitudes of the first intermediate level $\{m',M'\}$  and the initial one $\{M_1,M_1\}$. The cofactor $q_1$ quantifies the corresponding unconscious underestimation (overestimation) of the magnitude $M'$. The quantities $r_2$, $q_2$ and $r_3$, $q_3$ play the same roles in the transitions from the fist to second intermediate levels and from the second to final (third) levels. 

Turning to the given example, the logarithmic form of  the $N$-step relationship between the estimated value $m_2$ and the real magnitude $M_2$ is written as
\begin{equation}\label{FS:2}
\ln M_2 - \ln M_1  = \ln m_2 - \ln M_1 + \sum_{i=1}^N \ln q_i\,.
\end{equation}
This relationship has regular and random (irregular) components. First, there could be several configurations $\{r_1,r_2,r_3,\ldots\}$ leading from $M_1$ to $m_2$ different in the selected quantities and their number. 
Nevertheless if I introduce the value $\langle \ln r\rangle$ being a certain mean of the quantities $\{r_n\}$ the ``distance'' between  $\ln M_1$ and $\ln m_2$ can be quantified as $\langle \ln r \rangle\cdot \eta $, where the value $\eta$ not necessarily integer represents the number of steps $N$.   
Second, the cofactors $\{q_i\}$ are some random quantities reflecting the probabilistic nature of magnitude evaluation. Separating the mean value $\langle \ln q\rangle$ and random fluctuations $\delta\ln q_i$ with zero mean in $\ln q_i$ expression~\eqref{FS:2} can be approximately rewritten as 
\begin{subequations}\label{FS:3}
\begin{align}
\label{FS:3a}
\ln \left[\frac{m_2}{M_1}\right]  &= \langle \ln r \rangle \cdot \eta  \\
\label{FS:3b}
\ln \left[\frac{M_2}{M_1}\right] &= \big[\langle \ln r \rangle+ \langle \ln q\rangle\big] \eta + \epsilon \sqrt{\eta}\cdot\xi\,,
\end{align}
\end{subequations}
where $\xi$ is a random variable of unit amplitude and $\epsilon$ is a quantity determined by random variations of the underestimation cofactor, $\epsilon^2 \sim \langle (\delta\ln q_i)^2\rangle$. At the used level of rigor the parameters entering Exps.~\eqref{FS:3} should be regarded as given beforehand and constructing the specific relationship between them and the model of direct quantitative comparison of magnitudes deserves an individual investigation. 

In the case of small values of the parameter $\epsilon$  there is some critical number of intermediate steps of comparison, $N_c\sim \eta_c = 1/\epsilon^{2}$, such that if $N\ll N_c$ the random term in Exp.~\eqref{FS:3b} is ignorable and system~\eqref{FS:3} leads to the relationship
\begin{equation}\label{FS:4}
         m_2 = M_1^{1-\gamma} M_2^\gamma \,.
\end{equation}
where the exponent $\gamma = \langle \ln r \rangle/[\langle \ln r \rangle+ \langle \ln q\rangle]$ and $0<\gamma <1$. Expression~\eqref{FS:4} also holds for $M_2 \ll M_1/n_c^2$. I will call this limit case the \textit{power-law mode} of multi-step comparison. It capacity is bounded by the ratio $M_2/M_1$
\begin{equation*}
\ln \left[\frac{M_2}{M_1}\right]_c \sim \frac{\big[\langle \ln r \rangle+\omega \langle \ln q\rangle\big]}{\epsilon^2} \,.
\end{equation*}
For $\epsilon\sim 1$ this limit case does not exist. It should be noted that Exp.~\eqref{FS:4} may be continued into the region of direct commensurability and treated as a certain approximation. It exhibits the required underestimation or overestimation of test magnitudes and  the exponent $\gamma$ reflects the basic properties of image quantitative comparison. 

When the number of steps in the image comparison exceeds the critical value, $N\gg N_c$ the random term in Exp.~\eqref{FS:3b} become essential, $\epsilon\sqrt{\eta}\gtrsim 1$, which gives rise to unpredictable variations in the estimate 
\begin{equation*}
	         m_2 = \underbrace{\vphantom{M_2^\gamma}M_1^{1-\gamma} M_2^\gamma}_{\text{regular factor}}\ \times \!
	         \underbrace{\vphantom{M_2^\gamma}e^{-\epsilon\gamma\sqrt{\eta}\xi}}_{\text{random factor}}
\end{equation*}  
whose amplitudes can exceed the regular cofactor substantially. In this case the comparison of the \textit{absolute} values of magnitudes loses any sense.  The given problem can be overcome if in quantifying the images we \textit{consciously} remain at the ``logarithmic'' level where the random second term in Exp.~\eqref{FS:3b} is small in comparison with the first, regular one.  It gives the relationship
\begin{equation}\label{FS:5}
  \ln m_2 = (1-\gamma) \ln M_1 + \gamma \ln M_2
\end{equation}
for which the conversion to the level of absolute values is not allowed. I will call this case the \textit{logarithmic-law mode} of multi-step comparison.

It should be noted that laws~\eqref{FS:4} and \eqref{FS:5}, as it must be, are invariant with respect to changing the units of magnitudes, i.e., their form does not change within the replacement 
\[
(m_2,M_2,M_1)\rightarrow (m_2,M_2,M_1)\times C, \quad\text{where $C>0$ is an arbitrary constant.}
\]

\subsection{The level of units, sensory magnitudes, and the Fechner-Stevens dilemma}

Our life in the environment filled with various permanent stimuli should cause the emergence of their images in the mental space whose magnitudes are tacitly treated as the units in comparison with other images. So via referring to this \textit{level of units} it becomes possible to deal with mental evaluation of magnitudes on their own, i.e., without noting directly the image of comparison. This case is described by the expressions obtained before where the magnitude $M_1$ of the image $\mathbb{M}_1$ is set equal to unity, $M_1 = 1$, thereby all the other magnitudes are quantified in units of $M_1$.

The magnitude $M$ as a parameter of the $\mathcal{M}(m)$-cloud is a variable hidden from our mind because consciously we operate with these images as whole entities. However fixing the level of units we acquire the capacity to attribute \textit{consciously} some numerical measure to a given image just comparing it with the corresponding unit. In this way it becomes possible to speak a perceived ``intensity'' of stimulus---\textit{sensory magnitude}---$\mu(M|M_{\mathds{1}})$ being a certain function of the image magnitude $M$ provided the magnitude $M_{\mathds{1}}$ of the level of units is fixed, maybe, tacitly. Assuming that initially conscious estimates are converted with time into some automatic sensation of  external stimuli we may introduce the sensation magnitude $\mu(M)$ as a given on its own.

\begin{figure}
	\begin{center}
		\includegraphics[width=\columnwidth]{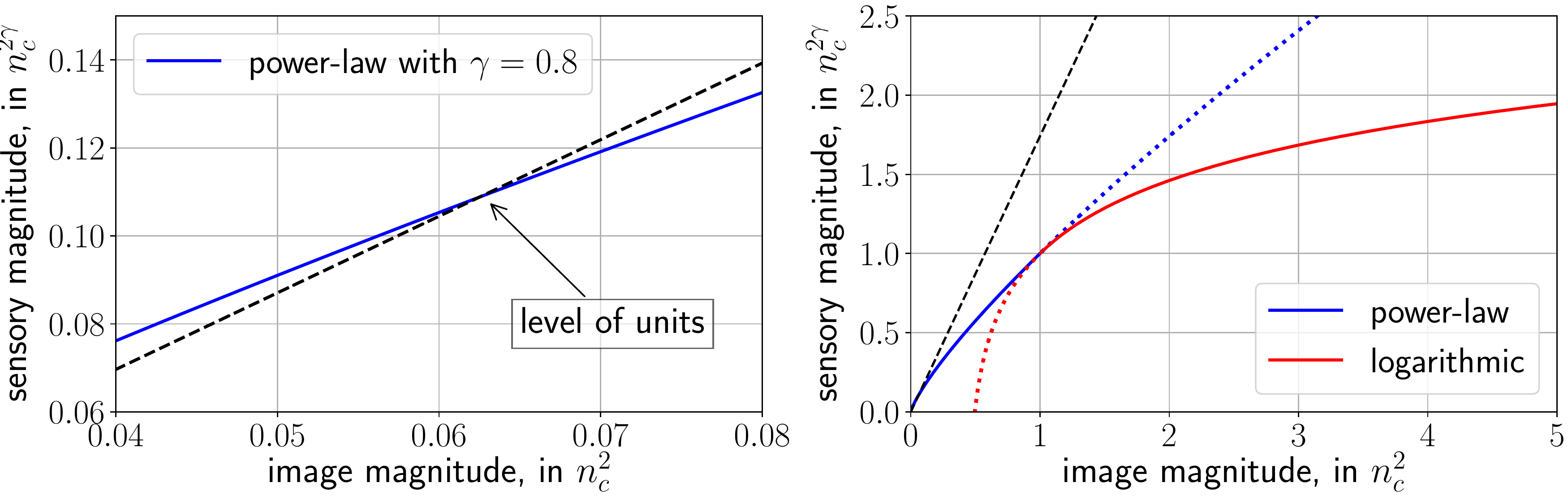}
	\end{center}
	\caption{Illustration of the relationship between the sensory magnitude and image magnitude constructed based on the quantitative comparison of images. Here dashed lines represents the image magnitude and in plotting $n_c=4$ and $\gamma =0.8$ were used.}
	\label{Fig:Fechstev}
\end{figure}

Keeping this possibility in mind I propose the following description of the \textit{sensation magnitude} within inner psychophysics.  After fixing the level of units we can divide the region of image magnitudes into three parts. First, it is the region of images whose magnitudes are too low to be recognized: 
\begin{multline*}
	\hspace*{2cm}\mathcal{R}_0:\quad M \lesssim \frac{1}{n_c^2}\,.\hfil
\end{multline*}
I call it the region of \textit{imperceptible images}. Here the sensory magnitude has to be set equal to zero, $\mu=0$. The region imperceptible images is followed by the region of \textit{commensurable images} specified by the condition 
\begin{multline*}
\hspace*{2cm}\mathcal{R}_c:\quad \frac{1}{n_c^2} \lesssim M\lesssim n_c^2\,, \quad\text{in Fig.~\ref{Fig:Fechstev} it is the region (0,1]}.\hfil
\end{multline*}
Its images are directly commensurable with the level of units. In this region it is natural to relate the sensory magnitude $\mu$ with the image magnitude $M$ using Exp.~\eqref{FS:4} as an approximating ansatz
\begin{equation}\label{FS:10}
\mu = M^\gamma\,, 
\end{equation}
which may be referred to as the inner Stevens' law. The last, third region
\begin{multline*}
\hspace*{2cm}\mathcal{R}_l:\quad M\gtrsim n_c^2\,, \quad\text{in Fig.~\ref{Fig:Fechstev} it is the region $[1,\infty)$}\hfil
\end{multline*}
is related to images admitting only multi-step comparison with the level of units. Here for simplicity I have accepted that $N_c\sim 1$, so the power-law mode of multi-step comparison does not exist. 

In this case we have to attribute image quantification to the ``logarithmic'' level and  expression~\eqref{FS:5} prompts us to regard $\ln m$ as the desired sensory magnitude. However, in the given form it is not relevant for two reasons. First, any rescaling of units leads to the anomalous transformation of such a sensory magnitude, instead of the standard renormalization in the form of some cofactor this sensory magnitude receives an additive component, $\ln m\rightarrow\ln m + C$. Second, the transition from the region $\mathcal{R}_c$ to $\mathcal{R}_l$ cannot be accompanied by a sharp drop in sensory magnitude. 

The two obstacles are overcome if the sensory magnitude $\mu$ is related to the image magnitude $M$ in the region $\mathcal{R}_l$, e.g., the ansatz 
\begin{equation}\label{FS:11}
\mu = \frac{n_c^{2\gamma}}{\ln\big[n_c^2(1-x_\text{cut})\big]}\cdot \ln\Big[M-x_\text{cut}n_c^2\Big].
\end{equation}
Its parameter  $x_\text{cut}=0.4$--0.6 for $n_c=4$--5 and $\gamma = 0.8$--1.0 endows the transition $\mathcal{R}_c\rightarrow\mathcal{R}_l$ with the continuity in the dependence $\mu(M)$ and its derivative. Expression~\eqref{FS:11} may be referred as the inner Fechner's law. 

Relating the image magnitude $M$ to the intensity $N$ of neural signals generated by the perceptual subsystem, e.g., via some proportionality $M\propto N$ and accepting the power-law transformation $N(S)$, Exp.~\eqref{M:StevPh}, we convert the constructed law of inner psychophysics into the corresponding combined Fechner-Stevens's law for outer psychophysics.

Summarizing the aforesaid, the introduction of sensory magnitude in the present way can be regarded as a solution to the Fechner-Stevens dilemma. Indeed the quantitative comparison of mental images of physical stimuli leads to Fechner's law and Stevens's law within inner psychophysics as particular asymptotics of the sensory magnitude dependence on the image magnitude in different limit cases (Fig.~\ref{Fig:Fechstev}). In this way the gist of the Fechner-Stevens dilemma is overcome  because in the presented explanation Fechner's law is not a consequence of Weber' law (Ekman's law in inner psychophysics).

\section{Conclusion: Inner psychophysics and physics of mind}    
 
The present paper for the first time has demonstrated  the relation between psychophysics studying quantitative regularities of external stimulus perception and a novel direction in physics---physics of mind---devoted to mathematical description of mental phenomena from the first-person point of view. 
The experimental data collected in the 20th century by psychophysics and neuroscience  of brain activity provide strong evidence for the existence of universal mechanisms responsible, in particular, for the main psychophysical laws. This universality opens a gate toward developing mathematical models for mental processes and phenomena turning to the notions we operate with in our mind.

As the main specific result, I have developed a phenomenological model for human evaluation of external stimuli that from a rather general standpoint explains the universal laws of inner psychophysics and proposed a solution to the Fechner-Stevens dilemma. 
Two premises take the central place in my constructions.
	
\textit{First}, mechanisms responsible for the main psychophysical laws should be attributed to the mental processes, conscious and unconscious. Actually I have accepted Fechner's point of view on the origin of regularities governing the relationship between external stimuli and their sensation, which prompted him to put forward the concept of \textit{inner psychophysics}. A number of arguments for the existence of a macrolevel mechanism common for all the sensory  modalities have been presented based on literature review of experimental data.

\textit{Second}, mental and cognitive phenomena admit a \textit{closed} mathematical description based on the notions and concepts that can be stated turning to the conscious experience as experienced from {the first-person point of view}. This premise is rooted in phenomenological description of human mind---the branch of philosophy launched by Husserl and Heidegger at the beginning of 20th century. Endowing phenomenology with mathematical formalism I came to the concept of two complementary components---objective and subjective ones---in  describing human behavior~\cite{ihor2017bookmind}. The objective component represents the physical world and is governed by classical laws of physics. The subjective component represents the inner world of humans and is assumed to be governed by its own laws not necessarily reducible to physical ones.

In the frameworks of psychophysics the objective component is the perceptual subsystem comprising our sense organs which directly convert external stimuli into neural signals. The subjective component is represented as a mental space with images of external stimuli whose emergence is caused by the neural signals generated by the perceptual subsystem. After emergence these images become individual entities of the mental space possessing their own properties and dynamics independent of the stimuli initially induced them.  In this sense the mental space is the macrolevel description of how neural signals generated by the perceptual subsystem are processed by the central subsystem affected by memory, the accumulated experience, etc.

Appealing to our causal experience I have postulated several properties these images have to exhibit with respect to their magnitude. In terms of subspace of their magnitudes: 
\begin{itemize}[topsep=0.15\baselineskip, itemsep=0.1\baselineskip,  parsep=0.15\baselineskip]
\item The image is a cloud in the magnitude subspace---a certain region with blurred boundary.
\item The structure of these clouds is universal for all images and is determined by the internal properties of mental space, external stimuli affect only the parameters of the cloud structure but not the structure itself. The image magnitude is a parameter specifying the mean cloud size.
\item The comparison of images with respect to their ``intensity'' is implemented based on the overlapping of their clouds which determines the probability of treating  the images as equivalent in magnitude.  
\item Two operations with images as whole entities is introduced in the mental space: (\textit{i}) the qualitative comparison of images with respect to their equality in magnitude and (\textit{ii}) the fusion of identical images whose number does not exceed a certain value. The quantitative comparison of image magnitudes is implemented based on the two operations.    
\end{itemize}
These features have enabled me to explain the main psychophysical regularities and to obtain plausibly estimates for the basic constants of inner psychophysics. In particular, it is:
\begin{itemize}
	\item The proportionality of just noticeable difference to image magnitude (Ekman's law) stemming from the qualitative comparison of images and the estimate of the proportionality fraction (Ekman's fraction).
	\item The universality of the maximal dynamic range of magnitude evaluation related to the direct commensurability of image magnitudes depending on magnitude ratio. Besides it is a plausible relationship between Ekman's fraction and the capacity of direct commensurability.
	\item The underestimation and overestimation of strong and weak stimuli as well as the hysteresis in the magnitude estimations in gradually increasing and then decreasing the intensity of external stimulus.  
\end{itemize}

Up to now there is a challenging problem in psychophysics, it is that fact that two basic laws---Fechner's law and Stevens' law---are in contradiction to each other. Each of them describes the human sensation of external stimuli but their functional forms  are different. Reconciling the two laws is a subject of long-term debates but no solution that would satisfy the participants has been found. In the present paper I have proposed an original solution to this Fechner-Stevens dilemma. Within the developed model for image comparison, Fechner's law and Stevens's law arise as asymptotics of one general law describing the relationship between sensory magnitude and external stimulus intensity in different limit cases. In this way the gist of the Fechner-Stevens dilemma is overcome  because in the presented explanation Fechner's law is not a consequence of Weber' law.

\end{document}